\documentclass[prb,twocolumn,amsmath,amssymb,floatfix]{revtex4-2}
\usepackage{graphicx}
\usepackage{bm}
\usepackage{placeins}
\usepackage[colorlinks=true,linkcolor=blue,citecolor=blue,urlcolor=blue]{hyperref}

\newcommand{\Md}{M_{\rm dice}}
\newcommand{\eps}{\varepsilon}

\begin{document}

\title{Quantum order by disorder among lozenge tilings:\texorpdfstring{\\}{ }
a staircase of hub-free crystals between the Lieb ferrimagnet and the antiferromagnet}

\author{Suhwan Yoo}
\author{Jeehoon Choi}
\author{Chaeyeon Kim}
\author{Hee Young Kwon}
\author{Changyeon Won}
\email{cywon@khu.ac.kr}
\affiliation{Department of Physics, Kyung Hee University, Seoul 02447, Republic of Korea}

\date{\today}

\begin{abstract}
Lozenge tilings of the triangular lattice define an exponentially large family of bipartite networks, from the dice lattice, a Lieb ferrimagnet with the largest sublattice imbalance, to square-type tilings, which are compensated antiferromagnets. For the classical Heisenberg antiferromagnet all of them are degenerate. We show that quantum fluctuations lift this degeneracy through a local mechanism: the zero-point energy is, to leading order, a sum of site terms that are convex in the coordination $z$, so that Jensen's inequality at fixed $\langle z\rangle=4$ selects the square-type tiling. An expansion about the Ising limit explains the convexity: a site with more bonds has more channels for quantum fluctuations, but each fluctuation frustrates more bonds. A magnetic field couples to the Lieb moment and competes with this selection. Because the moment is carried most cheaply by threefold and fivefold sites, sixfold hubs are avoided, and bond counting together with the Lieb--Mattis theorem limits hub-free tilings to $3/5$ of the dice moment. Quantum Monte Carlo and a complete enumeration of periodic tilings with small unit cells show that the classical jump is replaced by a staircase of periodic crystals. Its widest step is a hub-free crystal that saturates the bound; below it, threefold and fivefold defects pack ever more densely in a shallow sequence of crystals, controlled by a short-ranged attraction between defects of equal curvature and a repulsion between defects of opposite curvature. Selection energies scale as $JS$. In the height representation the coordination is the discrete Gaussian curvature and the Lieb moment its sublattice-staggered part. We give spectroscopic signatures and discuss realizations in nanographenes, atomic spin arrays and quantum simulators.
\end{abstract}

\maketitle

\begin{figure*}[t]
\includegraphics[width=0.9\textwidth]{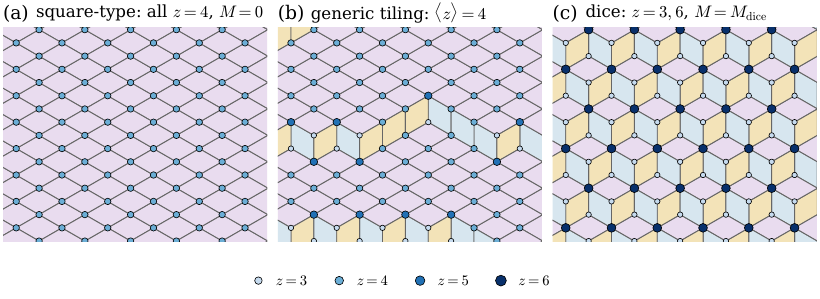}
\caption{Lozenge tilings of the triangular lattice and the bipartite networks they define. Rhombi are shaded by orientation, and sites are colored and sized by coordination $z$. (a)~Square-type tiling: all $z=4$, compensated ($M=0$). (b)~A generic tiling, with $z$ varying from site to site; every tiling has $\langle z\rangle=4$. (c)~Dice lattice: sixfold hubs and threefold rims, Lieb moment $\Md=SN/3$. All tilings have the same classical energy $-2JS^2N$.}
\label{fig:tilings}
\end{figure*}

\section{Introduction}

Order by disorder is the selection of a subset of classically degenerate ground states by fluctuations~\cite{Villain1980,Henley1989}. In frustrated magnets the degenerate manifold is usually a family of spin configurations on a \emph{fixed} lattice. Thermal or quantum fluctuations then pick the configurations with the softest excitation spectrum, most often collinear or coplanar ones. Harmonic zero-point energies often reduce to local effective Hamiltonians on the degenerate manifold~\cite{Henley2006}. In this paper we study a different kind of degeneracy, in which the degenerate object is the \emph{lattice itself}. The tilings below are fixed exchange graphs, each with its own Hamiltonian. Their ground-state energies select a lattice whenever the graph can rearrange while the spins stay in their ground state, i.e., when the tiling is an annealed degree of freedom. This is the situation in growth, self-assembly and programmable arrays, in which the network is a design or growth variable (Sec.~\ref{sec:disc}). We use ``order by disorder'' in this sense: quantum fluctuations select the graph. The tiling carries no kinetic term here, unlike in quantum dimer models; we compare the ground-state energies of fixed graphs, as in a Born--Oppenheimer separation in which the spins follow the graph adiabatically. The selection is therefore a zero-point stabilization of one lattice among degenerate ones.

Consider the lozenge (rhombus) tilings of the triangular lattice. Each tiling is obtained by pairing every up triangle with one of its down neighbors into a rhombus and deleting the shared edge. The remaining bonds form a planar bipartite graph in which every site has coordination $3\le z\le6$ and the mean coordination is exactly $\langle z\rangle=4$ [Fig.~\ref{fig:tilings}]. Two extreme members of this family are familiar. The dice ($\mathcal{T}_3$) lattice~\cite{Sutherland1986,Vidal1998} has sixfold hubs and threefold rims. The square-type tiling has only fourfold sites and is topologically the square lattice. By the Lieb--Mattis theorem~\cite{LiebMattis1962,Lieb1989}, the Heisenberg antiferromagnet on any of these graphs has a ground state with total spin $S_{\rm tot}=S|N_A-N_B|$. The dice lattice is therefore a ferrimagnet with the largest possible moment $\Md=SN/3$, while the square-type tiling is compensated. The family thus spans ferrimagnets and antiferromagnets continuously.

For classical spins every tiling has the same energy, $E_{\rm cl}=-JS^2N_b=-2JS^2N$, because every tiling has $N_b=2N$ bonds and collinear N\'eel order satisfies all of them. The number of tilings grows exponentially with $N$. Quantum mechanically the degeneracy is lifted, and two natural questions follow. Which lattice do quantum fluctuations select, and why? And how does a magnetic field, which couples directly to the Lieb moment and therefore favors the dice tiling, compete with that selection?

Quantum antiferromagnets on quasiperiodic and dice-derived networks have been studied by Jagannathan, Wessel and co-workers~\cite{Wessel2003,Jagannathan2006,Szallas2009,Jagannathan2012,Jagannathan2013}. Two of their results bear directly on the questions above. On the Penrose tiling, phason disorder lowers the ground-state energy~\cite{Szallas2009}. On the staggered dice lattice (SDL), a lozenge tiling with coordinations $z=3,4,5$ in equal numbers and $N_A=N_B$, local ``phason flips'' that create $z=3$ and $z=6$ sites raise the energy and enhance the staggered moment~\cite{Jagannathan2012}; domain walls between phason domains carry a positive line energy and feel a Casimir-like $1/d$ interaction~\cite{Jagannathan2013}. The authors described the enhancement of the staggered moment by geometric disorder as ``order by disorder''; we use the term for the selection of the graph itself. The two energy results have opposite signs. Section~\ref{sec:zero} shows that the SDL result follows from a site energy that is convex in the coordination, and that the Penrose result is consistent with it if disorder narrows the distribution of coordinations. These studies kept the two sublattices balanced, considered random or dilute flips, and did not include a field. The questions posed above involve sectors with different sublattice imbalance, systematic control of that imbalance, and the competition with a Zeeman field. To our knowledge they have not been addressed.

Our main results are as follows.
(i) The zero-point energy of any tiling is, to a good approximation, a sum of single-site energies $\eps(z)$, and $\eps(z)$ is convex. At zero field Jensen's inequality therefore selects the uniform $z=4$ tilings, all of which are square lattices as graphs. The resulting selection energy scales as $JS$ and can be computed for any tiling by counting coordinations [Fig.~\ref{fig:mech}].
(ii) Domain walls between dice and square regions have \emph{negative} energy [Fig.~\ref{fig:tilt}]. The field-driven crossover is therefore not a single first-order jump through phase coexistence.
(iii)~A second-order expansion about the Ising limit gives the selection energy as a sum over bonds of a convex function of $z_i+z_j$. It reproduces the convexity, the shape of the energy hull and the signs of the defect interactions without fit parameters.
(iv)~A tiling without sixfold hubs carries at most $\frac35\Md$. This bound follows from the Lieb--Mattis theorem and bond counting and is saturated only when every minority-sublattice site has $z=5$.
(v)~The crossover is a magnetization staircase of periodic crystals, established by enumeration of periodic tilings with up to 48 sites per cell and quantum Monte Carlo (QMC) [Figs.~\ref{fig:crystals} and~\ref{fig:stair}, Table~\ref{tab:crystals}]. Its deepest and widest step is a crystal that saturates the hub-free bound. Below it the steps are dense and narrow; the crystals are successive packing stages of $z=3$ and $z=5$ defects, governed by a short-ranged interaction in which like-sign curvature attracts [Fig.~\ref{fig:gas}].
(vi)~Each plateau carries distinct fluctuation, entanglement and dynamical signatures [Figs.~\ref{fig:fluct} and~\ref{fig:sqw}], and selection energies scale as $JS$, so that step fields expressed in $J$ depend only weakly on $S$ [Fig.~\ref{fig:spinS}].
(vii)~In the height representation, $z$ is exactly the discrete Gaussian curvature of a lifted surface, and the Lieb moment is its sublattice-staggered total, so that zero field selects folded, curvature-free surfaces and the field crumples them with conical defects (Appendix~\ref{app:geom}). In an effective tiling model the $\Md/2$ crystal melts at a temperature of a few percent of $J$, and local-flip dynamics become trapped on cooling (Appendix~\ref{app:finiteT}).

The paper is organized as follows. Section~\ref{sec:model} defines the model and its classical degeneracy, and Sec.~\ref{sec:methods} summarizes the methods. Section~\ref{sec:zero} explains the zero-field selection, its origin in the Ising limit and the hub-free bound. Section~\ref{sec:field} presents the field-driven crystals and the staircase. Sections~\ref{sec:fluct} and~\ref{sec:spinS} discuss fluctuations and the dependence on $S$. Section~\ref{sec:disc} discusses interpretation, platforms and limitations, and Sec.~\ref{sec:concl} concludes. The appendices give the geometric interpretation, tests of the models, numerical details and finite-size effects, and the finite-temperature behavior of an effective tiling model.

\section{Model and classical degeneracy}\label{sec:model}

We consider
\begin{equation}
H_{\mathcal T}=J\sum_{\langle ij\rangle\in\mathcal T}\mathbf S_i\cdot\mathbf S_j-h\sum_iS_i^z,\qquad J>0,
\label{eq:H}
\end{equation}
where the sum runs over the bonds of a lozenge tiling $\mathcal T$ of an $L\times L$ triangular torus ($N=L^2$ sites) with primitive vectors $\mathbf a_1$ and $\mathbf a_2$ at $60^\circ$. Each tiling is specified by a perfect matching of up and down triangles, which is equivalent to a dimer covering of the honeycomb lattice. Its height function~\cite{BloteHilhorst1982,Kenyon2006} has an average slope (tilt) that is conserved by the elementary hexagon flip, i.e., the rotation of three rhombi around a site with $z=3$. The tilt therefore labels topological sectors on the torus.

Every tiling has $2N$ bonds, so $\langle z\rangle=4$. The graph is bipartite with sublattices $A$ and $B$. By the Lieb--Mattis theorem~\cite{LiebMattis1962,Lieb1989} the ground state has $S_{\rm tot}=S|N_A-N_B|\equiv M$. We write $m=|N_A-N_B|/(2N)$ for the imbalance per site, so that $m=1/6$ for dice. The imbalance is not a topological invariant, since a single hexagon flip changes $N_A-N_B$ by 2. However, strips of dice and square-type tiling separated by straight walls provide a systematic family that interpolates between $m=0$ and $m=1/6$. We call it the ``tilt family'' below.

Classically, every tiling admits a collinear N\'eel state with energy $-JS^2$ per bond, so all tilings are degenerate at $h=0$. For $h>0$ the dice tiling wins immediately through its Zeeman energy $-hM$. The classical magnetization therefore jumps from 0 to $\Md$ at $h=0^+$ [dashed line in Fig.~\ref{fig:stair}(b)]. Everything that follows is a quantum effect.

\section{Methods}\label{sec:methods}

\emph{Quantum Monte Carlo.} All tilings considered here are bipartite, so Eq.~(\ref{eq:H}) is sign-free. We use the stochastic series expansion (SSE)~\cite{Sandvik1999} with directed loops~\cite{Syljuasen2002} and heat-bath exit probabilities, for general $S$ and with the field included in the bond operators. The code was checked against exact diagonalization (ED) of 27-site tori for $S=\frac12,1,\frac32$. The zero-field energy of a ferrimagnetic crystal is obtained from a run at a small field $h$ below its moment-raising gap $\Delta_+$ as $e_0=e+hm_{\rm meas}$, where $e$ is the measured energy per site including the Zeeman term. This is exact for the ground-state multiplet: on a fixed graph $S^z_{\rm tot}$ is conserved, the ground state is pinned at $S^z_{\rm tot}=M$ throughout a plateau, and all components of the Lieb multiplet have the same exchange energy, so $e_0$ does not depend on the field used. The field also gaps the former Goldstone mode, so moderate inverse temperatures suffice; we checked the convergence in $\beta$ for the references and for the deepest crystal. As a graph the square-type tiling \emph{is} the square lattice, so we use the thermodynamic-limit value $e_{\rm sq}$ of Ref.~\cite{Sandvik1997} as the zero-moment reference. This matters because finite-size corrections differ between the two kinds of tiling: they are sizable for the compensated tiling but negligible for crystals with an extensive moment, for which no size-dependent term proportional to the total moment arises. Moreover, an offset $\Delta e_{\rm sq}$ in $e_{\rm sq}$ changes every deviation $\delta e=e_0-e_{\rm lin}$ from the straight line $e_{\rm lin}(m)=e_{\rm sq}+6m\,(e_{\rm dice}-e_{\rm sq})$ by $\Delta e_{\rm sq}(1-6m)$, which is linear in $m$; it shifts only the first step and leaves the chords between ferrimagnetic crystals unchanged. Crystals whose cells do not fit an $L\times L$ torus are simulated on $n\times n$ copies of their cell (about $10^3$ sites), with periodic boundary conditions along $n\mathbf v_1$ and $n\mathbf v_2$. Temperatures, fields, convergence and autocorrelation checks, the scatter between independent seeds and the finite-size tests are given in Appendix~\ref{app:lswt}.

\emph{Spin-wave theory on arbitrary graphs.} For a bipartite graph with adjacency matrix $A$ and degree matrix $D$, linear spin-wave theory (LSWT)~\cite{HolsteinPrimakoff1940} gives magnon energies $\omega_\nu=JS|\lambda_\nu|$, where $\lambda_\nu$ are the eigenvalues of $\mathcal L^{1/2}P\mathcal L^{1/2}$, with $\mathcal L=D-A$ the graph Laplacian and $P={\rm diag}(\eta_i)$ the sublattice signs ($\eta_i=\pm1$). It also gives the ground-state energy $E=-JS^2N_b+\frac{JS}2(\sum_\nu|\lambda_\nu|-\sum_iz_i)$. We diagonalize the full bosonic Hamiltonian with Colpa's method~\cite{Colpa1978} to obtain spin deviations, Gaussian entanglement entropies and $S^{xx}(\mathbf q,\omega)$. Zero modes ($\omega<10^{-2}J$) are excluded.

\begin{figure}[t]
\includegraphics[width=\columnwidth]{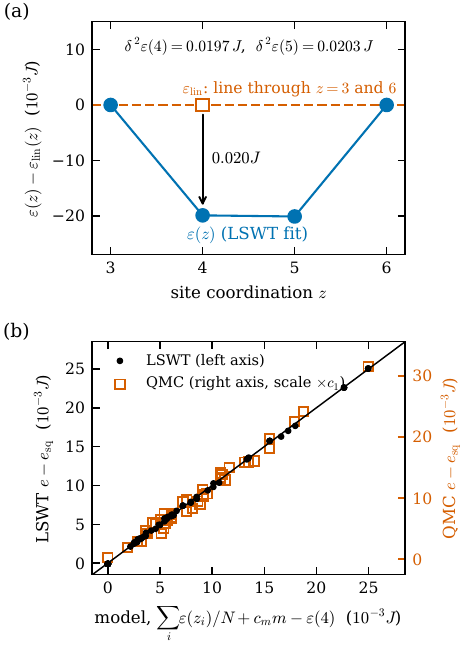}
\caption{Coordination mechanism of quantum tiling selection ($S=\frac12$). (a)~Single-site zero-point energies $\eps(z)$ from a fit of the LSWT energies of 75 tilings to Eq.~(\ref{eq:coord}), relative to the straight line $\eps_{\rm lin}(z)$ through $z=3$ and $z=6$ (dashed), on which mixtures of the two dice coordinations lie; subtracting it changes no energy difference. The coordination is an integer, and the segments are guides to the eye: convexity means that each point lies below the segment joining its neighbors, i.e., that the second differences $\delta^2\eps(z)=\eps(z+1)+\eps(z-1)-2\eps(z)$ are positive. The uniform $z=4$ tiling lies below the dice combination (open square) by $0.020J$ per site. (b)~Selection energies of many tilings, measured from the square-type tiling, vs the model energy. LSWT (dots) is read on the left axis and QMC (open squares) on the right axis, whose scale is stretched by the factor $c_1=1.27$ of the linear relation $e_{\rm QMC}=c_0+c_1e_{\rm model}$. On these axes both data sets fall close to the same straight line: the selection energy is close to linear in the model energy in both methods, and fluctuations beyond spin-wave theory enlarge it, by a factor that increases with defect density, without reordering the tilings beyond the level of arrangement energies. The rms residuals are $1.4\times10^{-4}J$ (LSWT) and $5$--$8\times10^{-4}J$ (QMC).}
\label{fig:mech}
\end{figure}

\emph{Role and validity of LSWT.} The conclusions of this paper rest on three kinds of evidence, which we keep apart. (i)~\emph{Exact statements:} the classical degeneracy, the ground-state spin $S_{\rm tot}=M$ (Lieb--Mattis theorem), the moment-charge identity, Eq.~(\ref{eq:charge}), the hub-free bound, Eq.~(\ref{eq:bound}), and the curvature identities of Appendix~\ref{app:geom}. (ii)~\emph{QMC:} every energy that enters the hull and the staircase, i.e., the two references, the crystals of Table~\ref{tab:crystals}, the tilt family and wall energies of Fig.~\ref{fig:tilt}, and the spin-$S$ data. (iii)~\emph{LSWT:} the screening of candidates before QMC, the fits of $\eps(z)$ and of the defect lattice gas, the interaction between walls, and the fluctuation, entanglement and spectral properties of Sec.~\ref{sec:fluct}. LSWT is used here as the leading quantum term of the $1/S$ expansion, of order $JS$, about a collinear state on an unfrustrated bipartite graph, for which the expansion has no cubic terms. Three checks support its use at $S=\frac12$. For the square lattice it gives the ground-state energy to within 2\% of the QMC value. The QMC energies of all tilings studied are close to linear in the LSWT-based model [Fig.~\ref{fig:mech}(b)]: QMC enlarges the selection energies by a factor that increases with defect density, and leaves their order unchanged except at the level of arrangement energies. And the QMC selection energies for $S=\frac12$, 1 and $\frac32$ extrapolate in $1/S$ to the LSWT values (Sec.~\ref{sec:spinS}). What LSWT cannot guarantee is the order of structures that differ only by arrangement energies; this limits the screening and is discussed in Sec.~\ref{sec:disc}.

\emph{Search over tilings.} We use three complementary searches. (a)~\emph{Periodic enumeration:} for supercells $\mathbf v_1=(p,0)$, $\mathbf v_2=(s,q)$ with $pq\le48$, we enumerate all tilings (up to $2\times10^7$ per cell) and keep the bipartite ones (on a torus some tilings are not bipartite). We group them into structural classes by a translation-invariant fingerprint (the coordination histogram together with the multiset of neighbor coordinations) and keep one representative per class, retaining every class within $0.03J$ of the coordination-model minimum at each $m$, so that tilings with the same coordination histogram but different arrangements are kept. Candidates within $2$--$3\times10^{-3}J$ of the coordination-model minimum (Sec.~\ref{sec:zero}) at each $m$, and every hub-free class for $pq\le36$, are evaluated with LSWT in the thermodynamic limit, and the hull structures with QMC. Supercells related by the point group $D_6$ are treated once (Sec.~\ref{sec:field}). (b)~\emph{Annealing:} simulated annealing in tiling space with local flips is performed at fixed tilt for $L=18,24,30$, first with the coordination model and then with the full LSWT energy. This produces the irregular networks shown in gray in Fig.~\ref{fig:stair}(a). (c)~\emph{Tilt family:} strips of dice and square tiling, used to extract wall energies.

\section{Zero field: why the square-type tiling wins}\label{sec:zero}

\subsection{A local, convex zero-point energy}\label{sec:local}

We first computed the LSWT energy of 75 tilings with widely different structure: tilt-family strips, and tilings with up to 200 random flips. Their energies are reproduced by a purely local form,
\begin{equation}
\frac{E_{\rm LSWT}}{N}\simeq\frac1N\sum_i\eps(z_i)+c_m\,m ,
\label{eq:coord}
\end{equation}
with a small positive coefficient $c_m$ that accounts for the weak cost of sublattice imbalance. Because $\sum_i(z_i-4)=0$ for every tiling, $\eps(z)$ is defined only up to a term $\lambda(z-4)$, which changes no energy. Only $\eps(4)$ and combinations that are invariant under this shift, such as the second differences and the costs of neutral groups below, are physical. The fitted values are listed in Appendix~\ref{app:tests}. The residual is about 2\% of the spread of energies among these tilings [Fig.~\ref{fig:mech}(b)].

The key property is that $\eps(z)$ is \emph{convex} [Fig.~\ref{fig:mech}(a)], a statement that is invariant under the shift above. Physically, the zero-point energy gained by spreading quantum fluctuations over more bonds saturates as $z$ grows, so replacing two $z=4$ sites by a $z=3$ and a $z=5$ site costs energy. Because $\sum_iz_i=4N$ is fixed for every tiling, Jensen's inequality gives
\begin{equation}
\frac1N\sum_i\eps(z_i)\;\ge\;\eps(\langle z\rangle)=\eps(4),
\label{eq:jensen}
\end{equation}
with equality only when all $z_i=4$. Every other tiling pays for its coordination variance. Tilings with all $z_i=4$ are not unique: besides the square-type tiling there are folded variants (Appendix~\ref{app:geom}). As graphs, however, all of them are square lattices, so the zero-field ground state is in every case the square-lattice antiferromagnet. The dice lattice, whose coordinations $3,3,6$ lie farthest from 4, is the most expensive at $h=0$. The same inequality gives the sign found in Ref.~\cite{Jagannathan2012}: flips in the SDL create $z=3$ and $z=6$ sites from $z=4$ and $z=5$ sites, which widens the coordination distribution and raises the energy. On the Penrose tiling, where phason disorder lowers the energy~\cite{Szallas2009}, the convex picture requires that disorder narrow the distribution of coordinations; we have not checked this.

\emph{How good is the local model?} We tested Eq.~(\ref{eq:coord}) in four ways (Appendix~\ref{app:tests}): by cross-validation on an enlarged training set, by adding bond-pair terms, out of sample on periodic crystals, and by comparing tilings with identical coordination histograms but different arrangements. The residuals show no systematic dependence on system size, number of flips or starting structure. Pair terms reduce them by about half, so correlations between neighboring coordinations are present but small. The arrangement energies at fixed histogram, described in Sec.~\ref{sec:gas}, are an order of magnitude smaller than the coordination energies that drive the selection. The model therefore captures the leading energy scale, such as the depth of the $\frac35\Md$ hull point, but not the finer arrangement energies. We use it only to screen candidates. Every energy on which a conclusion rests is computed with QMC.

QMC confirms this picture. For $S=\frac12$ the dice tiling lies above the square-type tiling by about 5\% of the exchange energy per site, whereas classically the two are degenerate. The QMC energies of all structures studied, including strips, periodic crystals and annealed networks, follow the coordination model after a single linear rescaling, $e_{\rm QMC}=c_0+c_1\,e_{\rm model}$ with $c_1>1$ [Fig.~\ref{fig:mech}(b)]: fluctuations beyond spin-wave theory enhance the selection energies but do not reorder them. A comparison at equal histogram but different imbalance, between the SDL and the $\Md/2$ crystal, shows that the cost of imbalance itself is larger in QMC than the rescaled model implies (Sec.~\ref{sec:field}). The coordination model is what makes the exhaustive searches of Sec.~\ref{sec:field} possible.

\subsection{Origin of the convexity: expansion about the Ising limit}\label{sec:ising}

The convexity of $\eps(z)$ can be derived. Split $H_{\mathcal T}$ at $h=0$ into $H_{zz}=J\sum S^z_iS^z_j$ and $H_\perp=\frac J2\sum(S^+_iS^-_j+{\rm h.c.})$, and treat $H_\perp$ as a perturbation of the N\'eel state, which is a ground state of $H_{zz}$ on every bipartite tiling. $H_\perp$ acts on one bond at a time: it lowers the spin at one end of $\langle ij\rangle$ and raises it at the other, with matrix element $JS$. The excited state costs $J[S(z_i+z_j)-1]$, because each of the other $z_i-1$ and $z_j-1$ bonds at $i$ and $j$ loses $JS$ and the bond itself loses $J(2S-1)$. To second order the energy per site is
\begin{equation}
e=-\frac{JS^2N_b}{N}-\frac{JS}{N}\sum_{\langle ij\rangle}\frac1{z_i+z_j-1/S}.
\label{eq:ising}
\end{equation}
Equation~(\ref{eq:ising}) has no adjustable parameter. It scales as $JS$, as the full selection energy does (Sec.~\ref{sec:spinS}), and it contains the qualitative content of this section and the next.

(1)~\emph{Convex site energies.} For defects that do not touch, every bond of a site with coordination $z$ has a $z=4$ partner. Using $\sum_i(z_i-4)=0$, Eq.~(\ref{eq:ising}) then reduces to the site form of Eq.~(\ref{eq:coord}) with
\begin{equation}
\eps(z)=-JS\,\frac{z}{z+4-1/S}+\lambda(z-4)+{\rm const}.
\label{eq:epsI}
\end{equation}
The fraction $z/(z+c)$ is concave, so $\eps(z)$ is convex. Coordination-dependent site energies were obtained earlier for Heisenberg star clusters on quasiperiodic tilings~\cite{Jagannathan2005}; Eq.~(\ref{eq:epsI}) gives their convexity in closed form. Physically, a site with more bonds has more channels for quantum fluctuations, but each fluctuation frustrates more bonds and costs more energy. The gain therefore saturates with $z$. For $S=\frac12$ the second differences of Eq.~(\ref{eq:epsI}) are between a third and a half of the LSWT values, and, per unit of moment charge (defined below), a hub with two $z=3$ sites costs about a third more than a $z=3$/$z=5$ pair; LSWT gives the same ordering (Sec.~\ref{sec:cost}). The values are listed in Appendix~\ref{app:tests}.

(2)~\emph{Contact energies.} When two defects share a bond, the bond term of Eq.~(\ref{eq:ising}) no longer separates into site terms. Replacing a $(3,4)$ and a $(4,5)$ bond by a $(3,5)$ and a $(4,4)$ bond costs $\frac J2(\frac15+\frac17-\frac13)>0$ for $S=\frac12$, so bonded defects of opposite curvature repel. Bonded like defects, $(3,3)$ and $(5,5)$, attract. The lattice-gas fit to LSWT in Sec.~\ref{sec:gas} finds the same signs and comparable magnitudes for the same contacts.

(3)~\emph{Shape of the hull.} Relative to its own straight line between the square-type and dice tilings, Eq.~(\ref{eq:ising}) reproduces the shape of the QMC hull of Sec.~\ref{sec:field}: the deviations of the hub-free crystals grow with their moment up to the $\frac35\Md$ crystal, where Eq.~(\ref{eq:ising}) also places the minimum, and decrease once hubs enter. It underestimates the overall scale, by up to about a quarter, both for these deviations and for $e_{\rm dice}-e_{\rm sq}$ (Appendix~\ref{app:tests}). Equation~(\ref{eq:ising}) is therefore a qualitative guide. All conclusions below rest on LSWT and QMC.

\subsection{The cost of a moment and the hub-free bound}\label{sec:cost}

Equation~(\ref{eq:coord}) also tells us how a tiling can acquire a moment most cheaply. On any bipartite tiling each bond joins $A$ to $B$, so $\sum_{i\in A}z_i=\sum_{i\in B}z_i=2N$. Writing $q_i=\pm(z_i-4)$, with $+$ on the minority sublattice $A$ and $-$ on $B$, this gives
\begin{equation}
m=\frac{N_B-N_A}{2N}=\frac1{8N}\sum_i q_i .
\label{eq:charge}
\end{equation}
Every site with $z\ne4$ thus carries a ``moment charge'' $q_i$. Because $\sum_i(z_i-4)=0$, charges come in neutral groups, and only the cost of a neutral group is defined unambiguously. The cheapest group is a $z=5$ site on $A$ with a $z=3$ site on $B$: charge $+2$ at cost $\eps(5)+\eps(3)-2\eps(4)$, which is positive by convexity. A hub requires a $z=6$ site on $A$ and two $z=3$ sites on $B$: charge $+4$ at cost $\eps(6)+2\eps(3)-3\eps(4)$. Per unit charge the hub is about 50\% more expensive in LSWT. To leading order, the energy therefore rises \emph{linearly} with $m$, with slope $4[\eps(5)+\eps(3)-2\eps(4)]+c_m$, as long as the moment can be carried by aligned $z=3$/$z=5$ pairs alone. It rises more steeply once hubs become unavoidable.

The largest moment that $z=3$/$z=5$ pairs can carry follows from bond counting. Let a tiling contain no hubs, so that $z_i\le5$ for every site. Then $2N=\sum_{i\in A}z_i\le5N_A$, i.e., $N_A\ge\frac25N$, and with $N_A+N_B=N$
\begin{equation}
m=\frac{N_B-N_A}{2N}\le\frac1{10},\qquad M\le\tfrac35\,\Md .
\label{eq:bound}
\end{equation}
By the Lieb--Mattis theorem this is a bound on the ground-state spin of every hub-free tiling. Equality requires $z_i=5$ on every site of $A$. The $B$ sites then have $\sum_{i\in B}z_i=2N$ on $N_B=\frac35N$ sites, i.e., $\langle z\rangle_B=\frac{10}3$, and $B$ sites may in principle have $z=3$, 4 or 5. The leading-order energy, Eq.~(\ref{eq:coord}), selects among these histograms. If $B$ carries only $z=3$ and $z=4$, two thirds of its sites have $z=3$ and one third $z=4$, i.e., $z_3{:}z_4{:}z_5=2{:}1{:}2$; a $z=5$ site on $B$ replaces a $z=4$ site and forces another $z=4$ site on $B$ to become $z=3$, which costs $\eps(3)+\eps(5)-2\eps(4)>0$ by convexity (the histogram $1{:}0{:}1$, for example, lies higher by $[\eps(3)+\eps(5)-2\eps(4)]/10$ per site). The leading-order optimum at the bound is therefore $2{:}1{:}2$. A larger moment requires hubs.

Whether the bound is reached by an actual tiling is a question of geometry. The enumeration of Sec.~\ref{sec:field} finds such a tiling in a 20-site cell. The leading-order energy $e_{\rm model}(m)$ is therefore a straight line from the square-type tiling to the hub-free $\frac35\Md$ crystal, followed by a steeper segment to dice. The $\frac35\Md$ crystal sits at the kink between them. This gives two predictions, both borne out below. The $\frac35\Md$ plateau is robust, because it rests on the leading term. Structures with $0<M<\frac35\Md$ are nearly degenerate at leading order, so the steps below it are decided by smaller arrangement-dependent energies, which we quantify in Sec.~\ref{sec:gas}.

\subsection{Tilt family and negative walls}

\begin{figure}[t]
\includegraphics[width=\columnwidth]{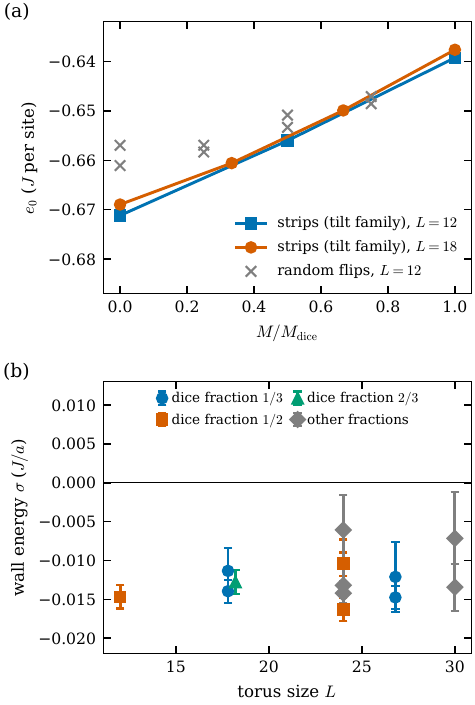}
\caption{(a)~QMC ground-state energies ($S=\frac12$) of the tilt family (parallel strips of dice and square tiling) for $L=12$ and $18$, together with tilings obtained by random hexagon flips (crosses). Random flips cost energy at fixed $M$; the classical energy is $-0.5J$ for all points. (b)~Energy per unit length of a dice$|$square wall extracted from the strips, for several dice fractions and torus sizes. All values are negative.}
\label{fig:tilt}
\end{figure}

Figure~\ref{fig:tilt}(a) shows the QMC energy of the tilt family as a function of its moment. The curve is slightly convex: intermediate strip patterns lie \emph{below} the straight line joining the square-type and dice tilings. Subtracting the bulk energies gives the energy per unit length of a dice$|$square wall, $\sigma$. It is negative for every width, dice fraction and system size studied [Fig.~\ref{fig:tilt}(b)]. The coordination picture explains the sign. A wall converts rim ($z=3$) and hub ($z=6$) sites of dice into $z=4$ and $z=5$ sites, which lowers $\sum_i\eps(z_i)$ because $\eps$ is convex. By contrast, random flips at fixed $M$ (crosses) always raise the energy, in agreement with Ref.~\cite{Jagannathan2012}. The sign of $\sigma$ is also opposite to that of the phason-flip walls inside the SDL, whose line energy is positive~\cite{Jagannathan2013}. Those walls separate two tilings with the same coordination histogram, so they cannot lower $\sum_i\eps(z_i)$; a dice$|$square wall can.

\begin{figure*}[t]
\includegraphics[width=\textwidth]{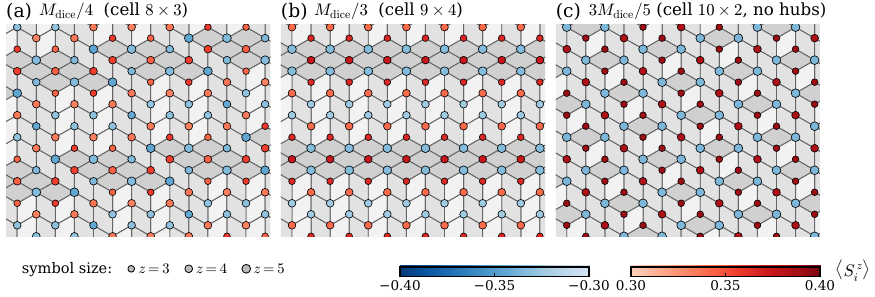}
\caption{Three periodic tilings of the lower convex hull of $e_0(m)$ (QMC, $S=\frac12$). (a)~$\Md/4$ crystal (cell $8\times3$, shear $s=2$): $z=3$/$z=5$ defects in a square-type background, each with one or two bonded neighbors of opposite curvature. (b)~$\Md/3$ crystal (cell $9\times4$, $s=4$): alternating $z=3$/$z=5$ chains. (c)~$\frac35\Md$ crystal (cell $10\times2$, $s=1$): the hub-free bound of Eq.~(\ref{eq:bound}), $z_3{:}z_4{:}z_5=2{:}1{:}2$, with $z=5$ on every minority site. Rhombi are shaded by orientation. Site color: local moment $\langle S^z_i\rangle$ in the field-polarized ground state (LSWT, thermodynamic limit), which does not change within a plateau. The hue gives its direction (red along $h$, blue opposite) and the shade its magnitude on the two scales shown, which start at $|\langle S^z_i\rangle|=0.30$: the moments are reduced most on the minority sublattice and least on the threefold majority sites. Symbol size: coordination $z$.}
\label{fig:crystals}
\end{figure*}

\begin{figure*}[t]
\includegraphics[width=\textwidth]{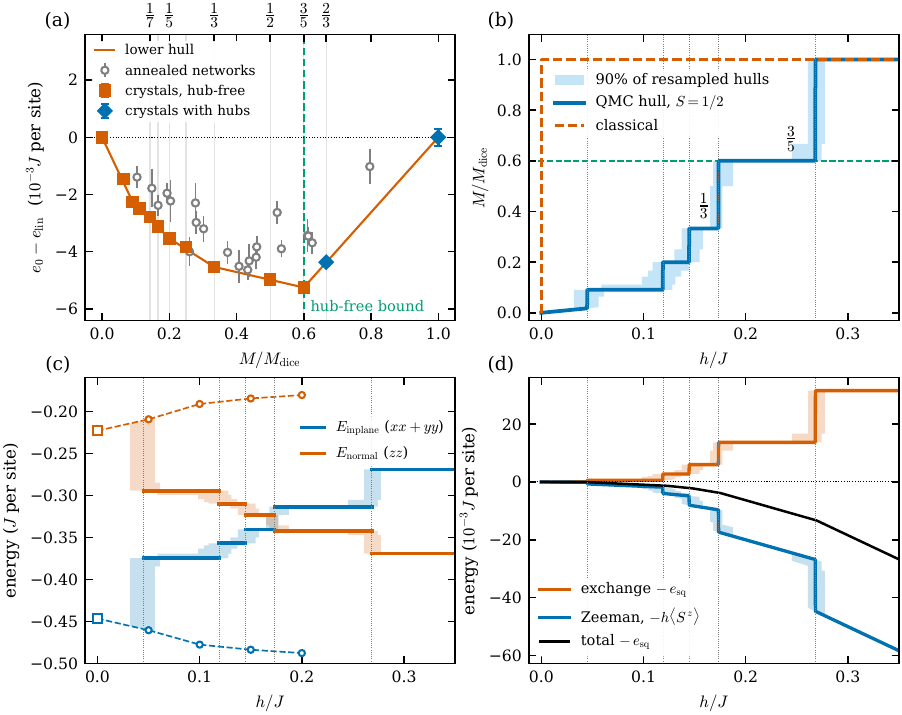}
\caption{Magnetization staircase ($S=\frac12$, QMC). (a)~Zero-field energy relative to the straight line between the square-type and dice tilings, $e_{\rm lin}(m)=e_{\rm sq}+6m\,(e_{\rm dice}-e_{\rm sq})$. Squares: best hub-free periodic crystal at each $M/\Md$ (Table~\ref{tab:crystals}); diamonds: crystals with hubs. Circles: irregular networks from LSWT annealing at $L=18,24,30$. The line is the lower convex hull, and the dashed vertical line marks the hub-free bound $M=\frac35\Md$, Eq.~(\ref{eq:bound}). (b)~$T=0$ magnetization obtained by minimizing $e_0-hm$ over the hull (solid; the square-type tiling contributes $m=\chi h$), compared with the classical jump at $h=0$ (dashed). Below $h\approx0.15J$ the individual steps are not resolved. Dotted vertical lines mark the step fields. In (b)--(d) the shaded bands contain 90\% of the curves obtained from 2000 hulls resampled within the QMC errors; they show how the uncertainty about which crystal is selected propagates to each quantity. (c)~In-plane ($xx+yy$) and normal ($zz$) exchange energy per site along the ground-state path (SSE, $\beta J=36$). Open symbols joined by dashed lines: square-type tiling, with the isotropic values $\frac23e_{\rm sq}$ and $\frac13e_{\rm sq}$ at $h=0$ (squares) and the spin-flop state at $L=24$, also outside its window (circles). (d)~Exchange energy relative to $e_{\rm sq}$, Zeeman energy and their sum along the same path.}
\label{fig:stair}
\end{figure*}

A negative wall energy has a direct consequence. When the field makes dice favorable, the system does not phase-separate into large dice and square domains. It instead gains energy by creating walls. In LSWT the energy per wall becomes more negative when the walls approach each other, down to the closest spacing compatible with bipartiteness (Appendix~\ref{app:tests}), so straight walls attract at short distance. The wall picture explains why intermediate structures win, but it does not select the crystal. Superlattices of straight walls lie well above the best crystals at the same $m$, by about half the depth of the hull. The reason is their histogram: straight-wall superlattices still contain hubs, whereas the crystals have none. The selection is controlled by the coordination histogram, as in Sec.~\ref{sec:cost}, and the negative wall energy is its consequence rather than an independent cause.

\section{Field-driven crystals and the magnetization staircase}\label{sec:field}

\begin{table*}[t]
\caption{Periodic crystals on or near the lower hull ($S=\frac12$). $M/\Md=6m$. Coordination counts are per cell, with cells $\mathbf v_1=(p,0)$, $\mathbf v_2=(s,q)$. $\delta e=e_0-e_{\rm lin}$ from QMC, per site, with $N$ the number of sites simulated ($\beta J=36$ on $L=36$ tori or $n\times n$ copies of the cell; dice at $L=\beta J=24$); the square-type reference is the thermodynamic-limit value~\cite{Sandvik1997}. The last columns give the field window in which the crystal is the $T=0$ ground state and the fraction of 4000 hulls, resampled within the QMC errors, in which it appears. Crystals without a window lie on or slightly above a chord of the hull, within about one standard error. Seed-to-seed spreads are $0.4$--$1.4$ times the binning errors.}
\label{tab:crystals}
\renewcommand{\arraystretch}{1.15}
\begin{ruledtabular}
\begin{tabular}{lcccccc}
$M/\Md$ & cell $(p,q,s)$ & $z_3{:}z_4{:}z_5{:}z_6$ & $\delta e$ ($10^{-3}J$) & $N$ & window ($h/J$) & present\\
\hline
0 & square & $0{:}1{:}0{:}0$ & 0 & $\infty$ & $<0.045$ & 1.00\\
$3/47$ & $(47,1,10)$ & $2{:}43{:}2{:}0$ & $-1.46(9)$\footnotemark[3] & 1175 & --- & 0.09\\
$1/11$ & $(11,3,4)$ & $2{:}29{:}2{:}0$ & $-2.25(10)$\footnotemark[4] & 825 & $0.045$--$0.119$ & 0.93\\
$1/9$ & $(27,1,4)$ & $2{:}23{:}2{:}0$ & $-2.49(9)$\footnotemark[4] & 972 & --- & 0.42\\
$1/7$ & $(14,3,3)$ & $4{:}34{:}4{:}0$ & $-2.79(15)$\footnotemark[1] & 1050 & --- & 0.19\\
$1/6$ & $(9,4,4)$ & $4{:}28{:}4{:}0$ & $-3.13(11)$\footnotemark[1] & 1296 & --- & 0.34\\
$1/5$ & $(10,3,8)$ & $4{:}22{:}4{:}0$ & $-3.54(12)$\footnotemark[1] & 1080 & $0.119$--$0.145$ & 0.72\\
$1/4$ & $(8,3,2)$ & $4{:}16{:}4{:}0$ & $-3.84(10)$\footnotemark[4] & 864 & --- & 0.28\\
$1/3$ & $(9,4,4)$ & $8{:}20{:}8{:}0$ & $-4.53(9)$\footnotemark[2] & 1296 & $0.145$--$0.173$ & 0.99\\
$1/2$ & $(6,4,4)$ & $8{:}8{:}8{:}0$ & $-4.98(9)$ & 1296 & --- & 0.48\\
$3/5$ & $(10,2,1)$ & $8{:}4{:}8{:}0$ & $-5.25(11)$\footnotemark[1] & 1280 & $0.173$--$0.268$ & 1.00\\
$2/3$ & $(9,4,7)$ & $16{:}6{:}12{:}2$ & $-4.37(12)$ & 1296 & --- & 0.48\\
1 & dice & $2{:}0{:}0{:}1$ & 0 & 576 & $>0.268$ & 1.00\\
\end{tabular}
\end{ruledtabular}
\footnotetext[1]{Mean of three independent runs (random seeds).}
\footnotetext[2]{Mean of two independent runs.}
\footnotetext[3]{Mean of four runs: three seeds at $h=0.04J$ and one at $h=0.02J$.}
\footnotetext[4]{Mean of five independent runs.}
\end{table*}

\subsection{Exhaustive search of periodic tilings}

Because the energy of a tiling is essentially a function of its coordination histogram, we can scan the complete set of periodic tilings with small unit cells. We took all supercells with up to 48 sites (403 cell shapes after reduction by the point group $D_6$), enumerated every tiling and retained the bipartite ones, keeping one representative per structural class (Sec.~\ref{sec:methods}). All cell shapes were enumerated completely; the largest cells contain up to about $10^8$ tilings each. Ranking by the coordination model is not sufficient for the final selection. It depends only on the histogram, whereas tilings with the same histogram differ by arrangement energies (Appendix~\ref{app:tests}). We therefore ranked about $2\times10^4$ candidates with LSWT in the thermodynamic limit (a Bloch calculation on the supercell, which agrees with $L=36$ tori for the crystals checked), and evaluated the hull structures with QMC.

The LSWT hull has points at $M/\Md=\frac3{47},\frac3{41},\frac1{11},\frac19,\frac17,\frac15,\frac14,\frac13,\frac35$ and 1. The $\frac3{47}\Md$ vertex, in a 47-site cell, is degenerate within the $k$-point accuracy with a $\frac1{15}\Md$ structure of the same defect family. We evaluated with QMC the crystals from $\frac3{47}$ to $\frac14$, the $\frac35\Md$ crystal, and the crystals at $\Md/6$, $\Md/3$, $\Md/2$ and $2\Md/3$, which have 24- and 36-site cells. The $\frac35\Md$ structure is the tiling that saturates the bound of Eq.~(\ref{eq:bound}): a 20-site cell $(10,2,1)$ with $z_3{:}z_4{:}z_5=8{:}4{:}8$, in which every minority site has $z=5$. In LSWT, too, the $\frac35\Md$ crystal is the deepest point of the hull, the $\Md/2$ crystal falls on the chord between $\frac13$ and $\frac35$, and the $\Md/6$ and $2\Md/3$ structures lie slightly above the hull.

Figure~\ref{fig:crystals} shows three crystals of the hull, and Fig.~\ref{fig:stair}(a) and Table~\ref{tab:crystals} give the QMC energies of all of them. Crystals whose cells do not fit an $L=36$ torus were simulated on $n\times n$ copies of their cells (Sec.~\ref{sec:methods}). The site colors in Fig.~\ref{fig:crystals} show that the field sorts the moment charges: every site with $z<4$ lies on the majority sublattice $B$ with its moment along $h$, and every site with $z>4$ lies on $A$ with its moment opposite to $h$, whereas a square-type tiling responds by a spin flop with a uniform $\langle S^z_i\rangle=\chi h$. Three features stand out.

(1)~The deepest point of the hull is the $\frac35\Md$ crystal (Table~\ref{tab:crystals}). It contains no hubs and saturates the hub-free bound. Its plateau is the widest of the staircase, and it appears in every resampled hull. This is the kink predicted in Sec.~\ref{sec:cost}: up to $\frac35\Md$ the moment is carried by $z=3$ and $z=5$ sites alone, and beyond it hubs are unavoidable.

(2)~The $\Md/2$ and $2\Md/3$ crystals lie on chords of the hull, the former on the chord between $\frac13$ and $\frac35$ and the latter on that between $\frac35$ and dice. Each is degenerate, within our resolution, with a coexistence of its neighbors, and each appears in about half of the resampled hulls. The $\Md/2$ crystal has the same coordination histogram as the compensated SDL of Refs.~\cite{Jagannathan2012,Jagannathan2013}. The difference is the arrangement: in the $\Md/2$ crystal the surplus sites are placed on one sublattice, giving $m=1/12$ instead of $0$. The comparison provides two checks. First, Eq.~(\ref{eq:coord}) predicts the LSWT energy of the SDL, $[\eps(3)+\eps(4)+\eps(5)]/3$, in agreement with the direct LSWT value of Ref.~\cite{Jagannathan2013}, and Jensen's inequality places its QMC energy~\cite{Jagannathan2012} above the square-lattice value, as found there and in our own SSE run (Appendix~\ref{app:lswt}). Second, because the two structures share a histogram, their difference isolates the cost of sublattice imbalance at fixed coordinations. This cost is more than twice as large in QMC as in LSWT, even after the rescaling of Sec.~\ref{sec:local}. It does not affect the hull, which is built from QMC energies, but it limits the use of the rescaled model for comparing structures at different $m$ with the same histogram.

(3)~Below $\Md/3$ the hull is dense. The crystals at $M/\Md=\frac1{11},\frac19,\frac17,\frac16,\frac15$ and $\frac14$, all hub-free arrays of $z=3$ and $z=5$ defects in a square-type background, lie close to the chords between their neighbors, within about the QMC errors. Only $\frac1{11}\Md$ and $\Md/5$ appear in most of the resampled hulls (Table~\ref{tab:crystals}), so the individual low-field plateaus are not resolved by our QMC. The $\Md/3$ crystal, a pattern of alternating $z=3$/$z=5$ chains, is robust. In LSWT the hull continues to lower moments, with points down to $\frac3{47}\Md$; in QMC the $\frac3{47}\Md$ crystal lies just above the chord from the square-type tiling to $\frac1{11}\Md$.

Irregular networks produced by annealing (gray circles) do not go below the hull. They come from about 160 annealing runs at $L=18$, 24 and 30, spanning five tilt sectors, fields up to $0.3J$ and several random seeds per setting. The best of them lie slightly above the hull, a few within one standard error of it. Within our resolution the ground state in a field is therefore a periodic crystal, although nearly degenerate networks exist.

\subsection{Staircase}

At $T=0$ the tiling selected by a field $h$ minimizes $e_0(m)-hm$ (for $S=\frac12$, where the Zeeman energy per site is $-hS_{\rm tot}/N=-hm$), with $-\frac12\chi h^2$ added for the compensated tiling. Minimizing over the hull yields the staircase of Fig.~\ref{fig:stair}(b):
\begin{equation*}
0\;\xrightarrow{\;h_c\;}\;\underbrace{\tfrac1{11},\ \dots,\ \tfrac15}_{\rm dense}\;\longrightarrow\;\tfrac13\;\longrightarrow\;\tfrac35\;\longrightarrow\;1,
\end{equation*}
with magnetization in units of $\Md$; the most probable sequence of the dense part and all step fields are listed in Table~\ref{tab:crystals}. The classical jump at $h=0^+$ is thus replaced by a staircase with two distinct parts. Above an onset field $h_c$ of about $0.05J$ (Sec.~\ref{sec:gas}) the magnetization rises to $\Md/5$ in a few narrow steps, each a denser packing of defects. Which intermediate crystals appear is not fixed by our data (feature 3). Above them come the $\Md/3$ plateau and then the $\frac35\Md$ plateau, which is about three times wider, before the jump to dice. This staircase is established within the searched space: periodic cells with up to 48 sites, plus the annealed networks. It is not a proof over all tilings. Each step corresponds to a first-order change of the \emph{lattice} and not only of the spin state.

Figures~\ref{fig:stair}(c) and \ref{fig:stair}(d) decompose the energy along this path. We measure the two parts of the exchange energy directly in SSE: $E_{\rm inplane}=-\langle n_{\rm od}\rangle/(\beta N)$ from the number of off-diagonal operators, and $E_{\rm normal}$ from the diagonal energy, which agrees with the equal-time estimator. Within a plateau both are independent of $h$. The exchange energy rises in steps, by $e_{\rm dice}-e_{\rm sq}$ in total, and the Zeeman energy $-hm$ pays for each step. The first step also reorients the order. In the square-type tiling the staggered moment lies perpendicular to $h$ (spin flop), so $E_{\rm normal}$ is a transverse correlation. In the ferrimagnetic crystals the moment is collinear with $h$, and $E_{\rm normal}$ is the longitudinal one, which is already markedly larger in magnitude at the first step. Along the staircase $|E_{\rm normal}|$ keeps growing and $|E_{\rm inplane}|$ shrinks, as the moments become more nearly classical. At $h=0$ the square-type ground state is a spin singlet with isotropic correlations, so that $E_{\rm normal}=e_{\rm sq}/3$.

\emph{Robustness.} We assess the staircase in three ways. (i)~\emph{Statistical errors.} Resampling all QMC energies within their errors (last column of Table~\ref{tab:crystals}) keeps the $\frac35\Md$ plateau and the jump to dice in every sample, with step fields that scatter by only a few percent [band in Fig.~\ref{fig:stair}(b)]. The $\Md/3$ plateau appears in almost every sample. The low-field crystals, and the $\Md/2$ and $2\Md/3$ crystals that lie on chords, appear in only part of the samples. The existence of the individual low-field plateaus is therefore not established, but the field range they occupy is. The scatter between independent seeds confirms the quoted errors (Appendix~\ref{app:lswt}). (ii)~\emph{System size.} The $\Md/2$ deviation agrees within errors between $L=24$ and $36$ and is consistent at $L=12$ (Table~\ref{tab:spinS}), the dice reference does not change between $L=18$ and $24$, and the square-type reference is taken in the thermodynamic limit (Sec.~\ref{sec:methods}). In LSWT the crystal energies at $L=36$, and on the $n\times n$ clusters, coincide with the thermodynamic limit. (iii)~\emph{Cell size.} The $\frac35\Md$ crystal has a 20-site cell and is the lowest tiling at the bound that we find in cells of up to 48 sites. The bound itself, Eq.~(\ref{eq:bound}), holds for all tilings, and the leading-order optimum at the bound is the histogram $2{:}1{:}2$. Larger cells could therefore lower the $\frac35\Md$ point only through arrangement energies, an order of magnitude below the leading term. They may still add hull points elsewhere, most likely at low moment, where the hull is dense.

The Lieb--Mattis theorem fixes $S_{\rm tot}=M$ at $h=0$. Because $S^z_{\rm tot}$ is conserved on a fixed graph and states with larger $S^z_{\rm tot}$ are separated by a magnon gap, the spin ground state stays pinned at $S^z_{\rm tot}=M$ throughout each plateau.

\subsection{A lattice gas of curvature defects}\label{sec:gas}

\begin{figure}[t]
\includegraphics[width=\columnwidth]{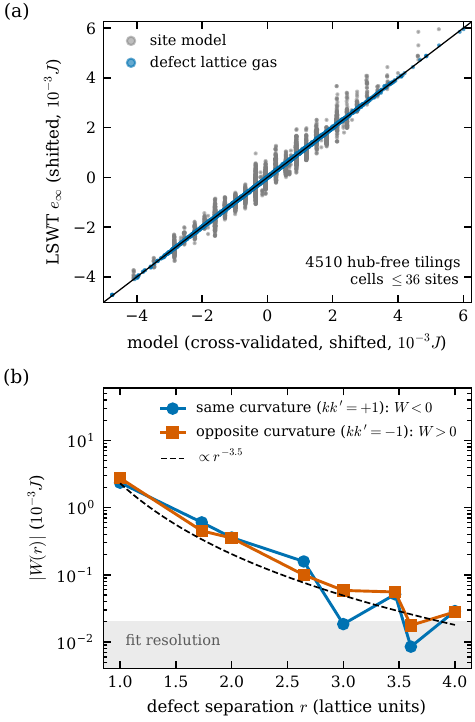}
\caption{Defect lattice gas, Eq.~(\ref{eq:gas}), fitted to thermodynamic-limit LSWT energies ($S=\frac12$) of all structurally distinct hub-free periodic tilings with cells of up to 36 sites. (a)~Cross-validated prediction vs LSWT: site model, Eq.~(\ref{eq:coord}), refitted to the same set (gray), and the lattice gas with pair terms up to $r=\sqrt7$ (blue). (b)~Magnitude of the pair energy $W$ for defects of the same (circles) and opposite (squares) curvature, averaged over the two sublattice relations; bonded nearest neighbors are included at $r=1$. Same-curvature pairs attract and opposite-curvature pairs repel. The dashed line, $\propto r^{-3.5}$, is a guide to the eye; the shaded band marks the resolution of the fit.}
\label{fig:gas}
\end{figure}

The site model treats every $z\ne4$ site independently and misses the arrangement energies that decide the low-field plateaus. We therefore describe a hub-free tiling as a lattice gas of curvature defects, the sites with $k_i=4-z_i=\pm1$, each on a sublattice $\eta_i=\pm1$, and fit
\begin{equation}
\frac{E_{\rm LSWT}}{N}=e_{\rm sq}+\eps_d\,\rho_d+c'_m\,m+\frac1N\sum_{i<j}W(r_{ij};k_ik_j,\eta_i\eta_j),
\label{eq:gas}
\end{equation}
where $\rho_d$ is the defect density. The pair energy $W$ depends on the distance and on whether the two defects have the same curvature and lie on the same sublattice, and bonded and unbonded nearest neighbors are distinguished. We fit Eq.~(\ref{eq:gas}) to the Bloch-LSWT energies of all 4510 structurally distinct hub-free tilings with cells of up to 36 sites, keeping pairs up to $r=\sqrt7$ [extending the cutoff to $r=4$, Fig.~\ref{fig:gas}(b), does not lower the residual]. The cross-validated residual is more than an order of magnitude smaller than for the site model [Fig.~\ref{fig:gas}(a)], and it stays as small on 48-site structures that were not used in the fit (Appendix~\ref{app:tests}). Three rules follow from the pair energies.

\emph{(1) Sign.} In 14 of the 16 pair channels ${\rm sign}\,W=-k_ik_j$, i.e., defects of the same curvature attract and defects of opposite curvature repel; the two exceptions are within the resolution of the fit. Bonded nearest neighbors dominate. The unlike contact, which we denote $W_1^{\rm unlike}$, is close to the Ising-expansion contact energy, and the like contact lies between the Ising values for $(3,3)$ and $(5,5)$ contacts (Sec.~\ref{sec:ising}). The field, which couples to the staggered curvature [Eq.~(\ref{eq:stagK})], is what sorts the curvature onto the two sublattices (Appendix~\ref{app:geom}); the pair energies then decide how the sorted defects are arranged.

\emph{(2) Range.} $|W|$ decreases by more than an order of magnitude between $r=1$ and $r=3$ [Fig.~\ref{fig:gas}(b)]. Beyond $r\approx3$ the couplings are below the resolution of the fit. The four resolved shells are compatible with a power law $\propto r^{-3.5}$, but two of them lie close to the resolution of the fit, so we do not attach significance to the exponent. Resolving $W$ into a curvature channel ($\propto k_ik_j$) and a moment channel ($\propto q_iq_j$, with $q_i=-\eta_ik_i$ the moment charge of Eq.~(\ref{eq:charge})) shows that the curvature channel dominates at short range, while at $r=2$ and $\sqrt7$ the moment channel is comparable and repulsive between equal moments. This is the sign of the magnon-mediated interaction between uncompensated moments, which is expected to decay as $r^{-3}$ for phason flips in the staggered dice lattice~\cite{Jagannathan2012}. In a polarized crystal all moments are equal, so this tail is repulsive and further favors evenly spaced dipoles.

\emph{(3) Packing.} Along the hull the defects pack more closely at each step. The number of bonded contacts with defects of opposite curvature, per defect, rises along the hull: 0 at $\frac1{11}\Md$ and $\frac19\Md$, 0.5 at $\frac17\Md$, 1.0 at $\Md/6$ and $\Md/5$, 1.5 at $\Md/4$, 2.0 at $\Md/3$ (alternating chains), 2.75 at $\Md/2$ and 3.0 at $\frac35\Md$. The number of unbonded nearest-neighbor contacts between like-curvature defects rises from 0 to 1.5. At the lowest densities many defects have no neighbor of opposite curvature (all of them at $\frac1{11}\Md$ and $\frac19\Md$, half at $\frac17\Md$, a quarter at $\Md/6$): dilute defects stay apart, as the repulsion between bonded unlike defects requires. At higher density unlike contacts become unavoidable, and the crystals combine them with attractive like contacts. In the $\frac35\Md$ crystal each $z=3$ site has three $z=5$ neighbors, and each $z=5$ site three $z=3$ and two $z=4$ neighbors. Each step field is set by the cost of the next contact divided by the moment it adds. Within each $m$, the model ranks the polarized structures in nearly the same order as LSWT and picks the same lowest structure.

The model also gives the onset of the staircase. Two field-aligned defects that do not touch cost $2\eps_d+c'_m/4$ and carry $mN=\frac14$, so the first step is expected near $h\approx8\eps_d+c'_m\approx0.06J$; a bonded pair would cost $W_1^{\rm unlike}$ more and set in at a higher field. The energy $\frac12\chi h^2$ that the square-type tiling gains by its spin flop shifts this estimate up only slightly. The LSWT hull indeed begins with dilute crystals of separated defects, with a first step close to this estimate, and the QMC onset is similar.

\emph{Dilute limit and the limits of finite cells.} A cell of $N_c$ sites cannot host a crystal with $0<m<1/(2N_c)$, so crystals more dilute than those enumerated, such as $\frac1{13}\Md$ or $\frac1{17}\Md$, are missing from our hull. Their energies relative to their neighbors would also be below the finite-cluster accuracy of LSWT and far below the QMC errors. The onset field, however, does not depend on them. Because $e_0(m)$ is convex, the first step leads from the square-type tiling to the crystal with the smallest $[e_0(m)-e_{\rm sq}]/m$, and this ratio decreases toward the dilute limit
\begin{equation}
h_c=\lim_{m\to0}\frac{e_0(m)-e_{\rm sq}}{m},
\label{eq:hc}
\end{equation}
which is the cost of one isolated, field-aligned $z=3$/$z=5$ pair divided by its moment and does not depend on which dilute crystals exist. Between $h_c$ and the field at which the most dilute evaluated crystal appears, the magnetization rises through all the missing crystals. The width of this range is set by the interaction between distant defects. A repulsive tail, such as the $r^{-3}$ magnon-mediated interaction between equal moments, makes every density appear and gives $m\propto(h-h_c)^{2/3}$, while an attractive tail makes the rise a single jump. In LSWT the ratio $[e_0(m)-e_{\rm sq}]/m$ is nearly constant from $\frac3{47}\Md$ to $\frac1{11}\Md$ and rises only above it (Appendix~\ref{app:lswt}): separated defects interact only weakly, consistent with the short range of $W$. In LSWT the missing dilute crystals are therefore confined to a narrow field range just above $h_c$, and the magnetization reaches $\frac1{11}\Md$ within a few thousandths of $J$ of $h_c$. Whether this rise is continuous or a jump cannot be decided at our resolution, and it does not affect the staircase at the scale of Fig.~\ref{fig:stair}. Applying the same saturation to the QMC energies of the dilute crystals gives $h_c\approx0.05J$, as in LSWT and consistent with the QMC step into $\frac1{11}\Md$ (Appendix~\ref{app:lswt}). Finally, because $W$ is short ranged, neighboring packing stages differ by contact numbers only, and their energies deviate from the chords between them by less than the QMC errors. This is why the low-field part of the hull is dense and shallow, and why our QMC resolves the field range of these steps but not each step individually.

\section{Fluctuations, entanglement and spectroscopic signatures}\label{sec:fluct}

\begin{figure}[t]
\includegraphics[width=\columnwidth]{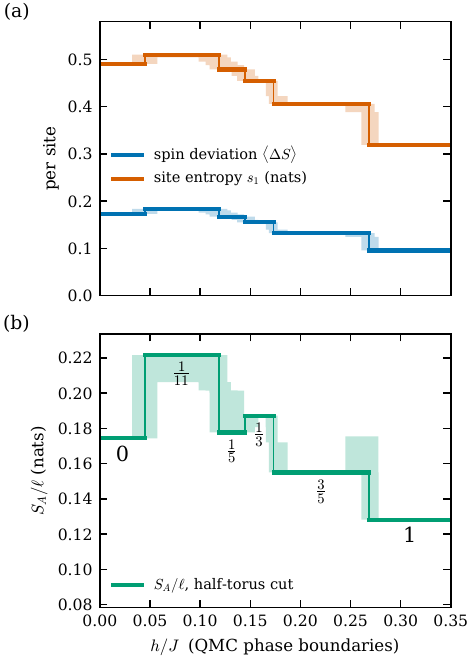}
\caption{Quantum-fluctuation measures of the LSWT ground state ($S=\frac12$) of the crystal selected at each field, using the QMC phase boundaries ($L=36$ tori, or $L=24$--$60$ tori for cells that do not fit $L=36$). (a)~Mean spin deviation $\langle\Delta S\rangle=\langle a^\dagger a\rangle$ and single-site entanglement entropy. (b)~Gaussian entanglement entropy of half the torus per unit boundary length, from the covariance matrices of the quadratures. Plateau labels give $M/\Md$. Shaded bands: 90\% of the hulls resampled within the QMC errors, as in Fig.~\ref{fig:stair}; the $\frac3{47}\Md$ crystal, which appears in 9\% of them, is not included.}
\label{fig:fluct}
\end{figure}

\begin{figure*}[t]
\includegraphics[width=\textwidth]{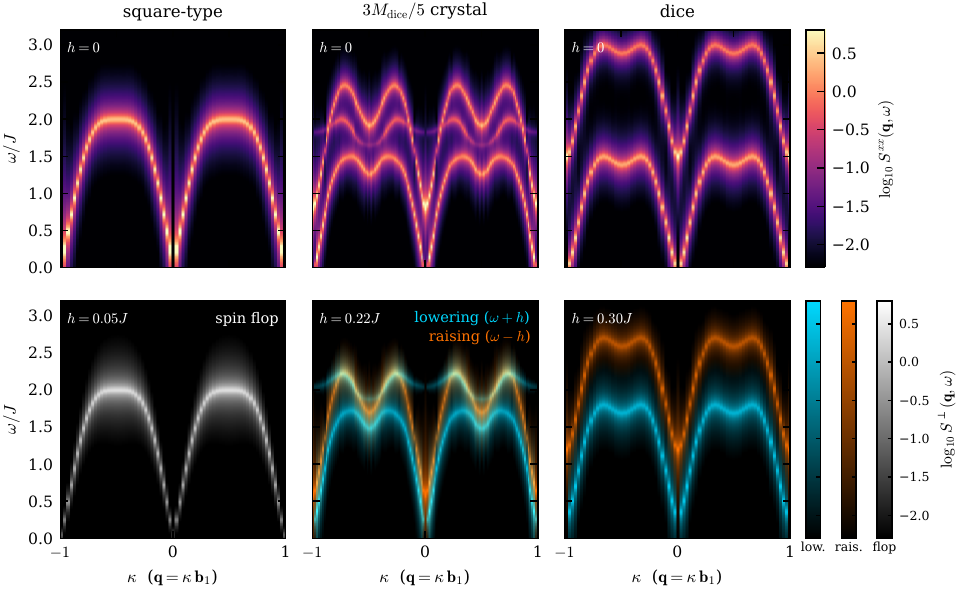}
\caption{LSWT transverse dynamical structure factor ($S=\frac12$, Lorentzian broadening $0.03J$, logarithmic scale) along $\mathbf q=\kappa\,\mathbf b_1$, where $\mathbf b_1$ is a reciprocal vector of the triangular lattice ($L=36$; $L=60$ for the $\frac35\Md$ crystal). Top row: $S^{xx}(\mathbf q,\omega)$ at $h=0$ for the square-type tiling (a single sharp magnon branch), the $\frac35\Md$ crystal (backfolded minibands and gaps set by its 20-site cell) and dice (a Goldstone branch, the dispersionless band near $1.5J$ that comes from the rim sites, and the hub branch above it). The frequency window contains the full bandwidth, $JSz_{\max}$. Bottom row: response transverse to a field inside each plateau. For the ferrimagnetic crystals the moment-lowering channel $\Delta S^z=-1$ (cyan, shifted to $\omega+h$) and the moment-raising channel $\Delta S^z=+1$ (orange, $\omega-h$) are shown separately. The square-type tiling is in the spin-flop state and shows a single unsplit branch (white). In the bottom row the hue identifies the channel and the brightness gives the intensity on the same logarithmic scale as in the top row (keys at right).}
\label{fig:sqw}
\end{figure*}

Because the spin state within a plateau does not change with $h$, the quantum-fluctuation content of the ground state changes only at the steps, where the lattice changes. Figure~\ref{fig:fluct} follows three measures along the staircase. The mean spin deviation rises slightly at the first steps and then decreases steadily toward dice, and the single-site entanglement entropy falls with it. The field thus reduces quantum fluctuations in discrete steps, by selecting lattices with more $z=3$ sites on the majority sublattice, whose moments are the most nearly classical. This is the mirror image of the zero-field selection, which favors the lattice with the largest zero-point gain. The half-system entanglement per unit boundary length is not monotonic. It rises at the first steps, where isolated defects add soft local modes, returns to the square-type value near $\Md/3$, and then falls below it toward dice. The values refer to finite tori with zero modes excluded, so small differences between crystals simulated at different $L$ are not significant.

The crystals can be distinguished spectroscopically [Fig.~\ref{fig:sqw}]. The square-type tiling shows a single linearly dispersing magnon branch. The dice tiling combines a Goldstone branch with the dispersionless band near $\omega\approx1.5J$ localized on the rims~\cite{Sutherland1986} and a moment-raising branch, carried mainly by the hubs, that starts at the same energy and extends to twice that value. The $\frac35\Md$ crystal shows backfolded minibands with small gaps at the zone boundaries of its 20-site cell. Its low-energy branch is square-like, and its upper spectral weight is spread over a band of nearly flat modes. In all three cases the top of the spectrum is $JSz_{\max}$, set by the largest coordination: the spectrum widens along the staircase as fivefold sites and then hubs appear. These features give a direct fingerprint of each plateau in inelastic neutron scattering or in magnon spectroscopy of engineered structures.

A field separates the two kinds of tiling (Fig.~\ref{fig:sqw}, bottom row). In a ferrimagnetic crystal the moment is locked along $h$, the Zeeman term commutes with $H_{\mathcal T}$, and every magnon carries $\Delta S^z=\pm1$. The spectrum is therefore split rigidly, $\omega\to\omega\mp h$, and the two sets of branches appear in different polarization channels. The Goldstone mode acquires the gap $h$. The moment-raising branch has an LSWT gap $\Delta_+$ that grows roughly as $8Jm$ along the staircase (Appendix~\ref{app:lswt}). A plateau is stable against magnon condensation as long as $h<\Delta_+$, i.e., $\Delta_+$ must exceed the upper edge of its window. This holds by a wide margin for every crystal on the path except $\frac1{11}\Md$, whose LSWT gap coincides with the upper edge of its QMC window; for this crystal the nature of the upper edge, a change of lattice or a magnon condensation, is not determined. Within its window each crystal therefore has zero differential susceptibility, and each step is a change of lattice, not a magnon condensation. The square-type tiling responds instead by a spin flop. Its magnons are not eigenstates of $S^z$, and the uniform mode acquires the gap $h$, so that $\chi$ remains finite. In a field the two kinds of tiling therefore differ in symmetry. The spin-flop state breaks the $U(1)$ symmetry about $h$ and keeps one linearly dispersing Goldstone mode. A ferrimagnetic crystal is collinear with $h$, preserves this symmetry, and has no gapless mode: its former Goldstone mode is gapped by $h$. In the limit $h\to0$ the difference becomes one of Goldstone-mode type. In the compensated tiling the two broken generators have a vanishing commutator, $\langle[S^x_{\rm tot},S^y_{\rm tot}]\rangle=i\langle S^z_{\rm tot}\rangle=0$, and give two linearly dispersing (type-A) modes. In a ferrimagnetic crystal the commutator is $iM\neq0$, the two generators form a canonical pair, and a single quadratically dispersing (type-B) mode remains~\cite{NielsenChadha1976,WatanabeBrauner2011,Watanabe2012,Hidaka2013}; the stability of such ferrimagnetic order, which has no tower of states, is discussed in Ref.~\cite{Rademaker2020}.

\section{Dependence on spin length}\label{sec:spinS}

\begin{figure}[t]
\includegraphics[width=\columnwidth]{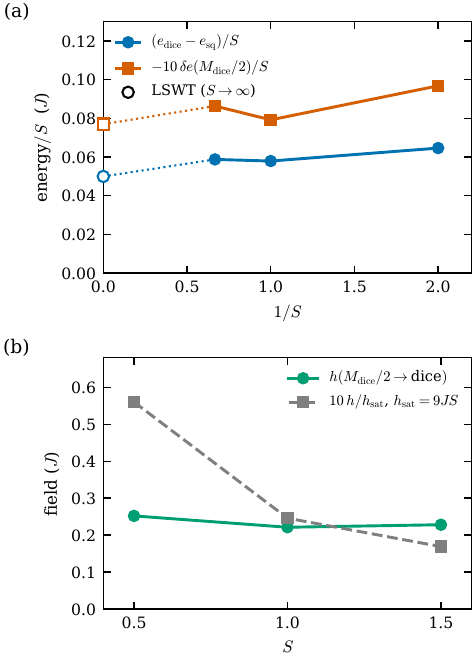}
\caption{Spin-$S$ QMC ($L=12$, $\beta JS=12$). (a)~Selection energies per unit $S$: dice minus square-type (circles) and the $\Md/2$ deviation (squares, $\times(-10)$). Open symbols: LSWT ($S\to\infty$); dotted lines join them to the QMC value at the largest $S$ as guides to the eye. Both approach finite limits, i.e., selection energies scale as $JS$. (b)~The field at which the $\Md/2$ crystal gives way to dice at this system size (circles) depends only weakly on $S$ in units of $J$. Relative to the saturation scale, the intermediate region therefore shrinks as $1/S$ (squares).}
\label{fig:spinS}
\end{figure}

\begin{table}[b]
\caption{Spin-$S$ dependence (QMC, $L=12$). Energies per site in units of $J$.}
\label{tab:spinS}
\begin{ruledtabular}
\begin{tabular}{lccc}
$S$ & $(e_{\rm dice}-e_{\rm sq})/S$ & $\delta e(\Md/2)/S$ & $h(\Md/2\to{\rm dice})$\\
\hline
$\frac12$ & 0.065 & $-0.0097$ & 0.25\\
1 & 0.058 & $-0.0079$ & 0.22\\
$\frac32$ & 0.059 & $-0.0086$ & 0.23\\
LSWT & 0.050 & $-0.0077$ & ---\\
\end{tabular}
\end{ruledtabular}
\end{table}

Classically the selection energy vanishes, and in spin-wave theory it is of order $JS$, whereas the exchange energy is of order $JS^2$. QMC for $S=\frac12$, $1$ and $\frac32$ confirms this scaling (Fig.~\ref{fig:spinS} and Table~\ref{tab:spinS}). The ratios $(e_{\rm dice}-e_{\rm sq})/S$ and $\delta e(\Md/2)/S$ are nearly constant and of the same size as the LSWT values ($S\to\infty$). (The $\frac35\Md$ crystal does not fit an $L=12$ torus, so we use the $\Md/2$ crystal, which lies on the hull chord next to it, as the representative hub-free crystal.) The same $JS$ scaling follows from Eq.~(\ref{eq:ising}), whose bond terms depend on $S$ only through the shift $1/S$ in the denominator. Both the energy differences and the Zeeman energy $hM\propto hS$ are linear in $S$, so the step fields are expected to be $S$-independent in units of $J$. We have checked this for one step: the field at which the $\Md/2$ crystal gives way to dice is nearly the same for all three spins (Table~\ref{tab:spinS}). Measured against the saturation field, which grows as $JS$ ($8JS$ for the square lattice and $9JS$ for dice in LSWT), the intermediate region therefore shrinks as $1/S$. The staircase is a quantum effect, most pronounced for $S=\frac12$.

The same holds for the $\frac35\Md$ crystal (Appendix~\ref{app:lswt}). The selection energies per unit $S$ and the field of the step from $\frac35\Md$ to dice all decrease by 10--15\% from $S=\frac12$ to $\frac32$, consistent with a $1/S$ correction, and a linear extrapolation in $1/S$ reproduces the LSWT values. Selection energies are therefore of order $JS$ with a correction of relative order $1/S$.

\section{Discussion}\label{sec:disc}

\emph{Height-model interpretation.} Appendix~\ref{app:geom} makes the height representation concrete: Eq.~(\ref{eq:coord}) is an effective height model whose potential is a convex function of the local Gaussian curvature of the lifted surface. The field acts as a chemical potential for the sublattice imbalance. In this language the plateaus are \emph{locked} phases of the height model, commensurate crystals analogous to the ordered phases of quantum dimer models~\cite{Rokhsar1988,Moessner2001}, in particular of the hexagonal-lattice dimer model, whose dimer coverings are the lozenge tilings~\cite{MoessnerSondhiChandra2001}. The staircase is analogous to the devil's staircases of competing-interaction systems~\cite{Bak1982}, but here the underlying potential is set by the convexity of $\eps(z)$ and not by an imposed long-range interaction. Our data establish a robust skeleton, the $\frac35\Md$ plateau, the $\Md/3$ plateau and the jump to dice, together with a dense set of narrow steps at low field. In the low-field part the energies of successive packing stages lie close to the chords between them. Near the onset, the dilute crystals that finite cells cannot host are confined to a narrow field range just above $h_c$ (Sec.~\ref{sec:gas}); whether they form a complete devil's staircase there, as a repulsive long-range tail would imply, is open but immaterial at the scale of the observable staircase.

\emph{Relation to known plateaus.} Magnetization plateaus in quantum magnets usually arise from a crystal of magnetic excitations on a \emph{fixed} lattice~\cite{Kageyama1999,Honecker2004,KarlovaStrecka2026}. Plateaus can also be stabilized by a coupling to the lattice, as in the half-magnetization plateau of the pyrochlore antiferromagnet, where a structural distortion modulates the exchange constants~\cite{PencShannonShiba2004}, and in frustrated spin chains coupled to adiabatic phonons~\cite{Vekua2006}. In that mechanism the bond network keeps its topology and only the bond strengths change. Here the plateau is a crystal of the lattice itself: the topology of the network changes from step to step, all bonds keep the same $J$, and the spin state on each plateau is the Lieb--Mattis ground state of that network. The mechanisms could coexist if bond strengths are also varied. Studying that would require going beyond the uniform-$J$ model~(\ref{eq:H}).

\emph{Platforms.} Three classes of system make the graph of exchange bonds a design or growth variable. (i)~\emph{$\pi$-magnetic nanographenes}, whose ground-state spin is set by the bipartite network of $\pi$ orbitals through sublattice imbalance and topological frustration, and whose building blocks can be fused into extended chains and networks~\cite{Mishra2020,Mishra2021}. (ii)~\emph{Atom-by-atom spin arrays} on surfaces, which realize arbitrary two-dimensional coupling graphs and allow local spectroscopy of spin excitations~\cite{Khajetoorians2019}. (iii)~\emph{Programmable quantum simulators}, such as Rydberg arrays, in which the connectivity is reconfigurable and antiferromagnetic correlations on hundreds of sites have been measured~\cite{Scholl2021}. These experiments realize Ising-type antiferromagnets, so a Heisenberg version of Eq.~(\ref{eq:H}) would require a different encoding. In (ii), and in (iii) once Heisenberg couplings are encoded, the energies $e_0(m)$ of different tilings can be compared directly, and the $\frac35\Md$ crystal, with a 20-site cell, is a concrete target structure. In these platforms the network is fixed by synthesis or by programming rather than annealed, so the staircase is best read as an energetic guide for designing the lowest-energy network at a given field, not as a sequence of transitions in one sample. In a system whose tiling can relax by local flips, a field sweep would in addition be limited by kinetics (Appendix~\ref{app:finiteT}) and by the conservation of the tilt: the $\frac35\Md$ crystal lies in a tilt sector different from those of the square-type and dice tilings, so reaching it requires a global rearrangement or free boundaries. In (i) the step fields, of order $0.05$--$0.3J$, and the melting temperature of the $\Md/2$ crystal in the effective model, a few percent of $J$, give the relevant scales once $J$ is known. In small systems ($N\sim10^2$) a further scale enters: a few uncompensated defects in a compensated background carry a size-dependent energy $\alpha|\mu|/N$, with $\alpha$ of a few $J$ for the staggered dice lattice in our LSWT, ED and QMC checks, related to the low-energy states of locally uncompensated regions in diluted antiferromagnets~\cite{WangSandvik2006}. This is comparable to the step fields at $N\sim10^2$ and would shift the lowest steps in finite arrays.

\emph{Limitations.} (a)~The enumeration is exhaustive for cells of up to 48 sites (LSWT screening), and the hull structures were evaluated with QMC on clusters of about $10^3$ sites. Larger cells may add steps or shift boundaries, most likely at low field. In particular, crystals more dilute than $\frac3{47}\Md$ cannot be hosted by our cells; the dilute-limit argument of Sec.~\ref{sec:gas} confines them to a narrow range just above the onset field $h_c$, which itself is a property of isolated defect pairs and not of the cells. The hub-free bound and its saturation by the $\frac35\Md$ crystal do not depend on cell size. (b)~Below $\Md/3$, and for the $\Md/2$ and $2\Md/3$ crystals, the margins are comparable to the QMC errors. The low-field crystals were simulated with two to five independent seeds, the $\Md/2$ and $2\Md/3$ crystals with one. That the deepest crystal has a cell incompatible with an $L=36$ torus shows that conclusions at the level of the arrangement energies depend on the completeness of the search. (c)~The coordination model is used only for screening. The defect lattice gas is far more accurate, but it is fitted to LSWT, which underestimates the QMC deviations by a density-dependent factor, and it covers only hub-free tilings. (d)~The finite-temperature results refer to an effective tiling model at fixed tilt without thermal spin fluctuations. The quoted melting temperatures are scales of that model, and the non-recrystallization is kinetic trapping. (e)~Irregular networks were generated with a model-assisted annealing that is not exhaustive. No irregular network we evaluated with QMC lies below the periodic hull, but a few lie within one standard error of it. (f)~We checked the $\beta$ convergence for the square-type reference and the $\frac35\Md$ crystal only (Appendix~\ref{app:lswt}); the other crystals are assumed to be converged, which is consistent with their gaps. Each crystal was simulated at a single system size, except the $\Md/2$ crystal. QMC was applied to the best LSWT structure at each $m$, not to the next few candidates, so an LSWT ordering error at the level of the arrangement energies would go undetected; Appendix~\ref{app:tests} shows that such errors are rare but not absent. (g)~The model contains only uniform nearest-neighbor exchange. Random, environment-independent variations of $J$ or further-neighbor couplings comparable to the arrangement energies could reorder the low-field crystals. A systematic dependence of $J$ on the coordination is much less harmful, as estimated in the next paragraph.

\emph{Coordination-dependent exchange.} In a real network the bond strength may depend on the local environment. Let $J_{ij}=J[1+\gamma(z_i+z_j-8)]$, so that bonds between highly coordinated sites are stronger for $\gamma>0$. To first order in $\gamma$, approximating the bond correlation by its mean value $c=\langle\mathbf S_i\cdot\mathbf S_j\rangle\simeq e_{\rm sq}/2J=-0.335$ and using $\sum_{\langle ij\rangle}(z_i+z_j)=\sum_iz_i^2$, the energy per site changes by
\begin{equation}
\delta e=\gamma Jc\,\big(\langle z^2\rangle-16\big)=\gamma Jc\,{\rm Var}(z).
\label{eq:gamma}
\end{equation}
For every perfectly sorted hub-free crystal ${\rm Var}(z)=8m$, because its histogram is fixed by $m$ (Sec.~\ref{sec:cost}). Equation~(\ref{eq:gamma}) is then linear in $m$: it shifts all steps between hub-free crystals by the same field, $\delta h=8c\gamma J$, and leaves their sequence, and the low-field structure, unchanged to this order. Only hubs break the linearity. Dice has ${\rm Var}(z)=2$ instead of $8m=\frac43$, so the step into dice moves by an additional field of the same order, and the $\frac35\Md$ plateau narrows; it closes only for $\gamma$ of a few percent, and the zero-field preference for the square-type tiling over dice is reversed only at a slightly larger $\gamma$. The central results are therefore stable against coordination-dependent exchange of a few percent; bond-to-bond variations of $c$ beyond the mean-value approximation enter at the level of the arrangement energies.

\section{Conclusion}\label{sec:concl}

We have shown that the classically degenerate family of lozenge-tiling networks is organized, quantum mechanically, by one rule: the zero-point energy is a convex function of the site coordination. An expansion about the Ising limit shows where the rule comes from: a site with more bonds has more channels for quantum fluctuations, but each fluctuation frustrates more bonds, so the gain saturates. This rule selects the uniform square-type tiling at zero field, gives dice$|$square-type walls negative energy, and, in competition with the Zeeman energy of the Lieb moment, stabilizes, within the searched space, a staircase of periodic crystals rather than the classical jump or a disordered network. Because a moment is carried most cheaply by $z=3$ and $z=5$ sites, hubs are avoided as long as possible, and bond counting together with Lieb's theorem limits hub-free tilings to $M\le\frac35\Md$. A crystal with a 20-site cell, in which every minority site is fivefold, saturates this bound. It is the deepest point of the energy hull and the central object of this work, stable over the widest field window of the staircase, a window set by selection energies that scale as $JS$. Below it the staircase is dense: crystals of $z=3$ and $z=5$ defects follow one another in narrow windows, held in place by a short-ranged attraction between curvature defects of the same sign and a repulsion between defects of opposite sign. Geometrically, zero field selects folded, curvature-free surfaces, and the field crumples them with sorted conical defects, because the Lieb moment is exactly the sublattice-staggered Gaussian curvature; the $\frac35\Md$ crystal is the densest packing of sublattice-sorted $\pm\pi/2$ cones and saddles. These results identify quantum order by disorder acting on the geometry of the exchange network itself. They also provide concrete target structures and energy scales for nanographene networks, atom-by-atom spin arrays and quantum simulators, in which that geometry can be designed.

\section*{Data availability}
The data that support the findings of this article are openly available~\cite{DataRepo}. The repository contains the SSE QMC, LSWT and enumeration codes, the crystal tilings, the QMC logs and the figure scripts.

\begin{acknowledgments}
This research was supported by the National Research Foundation (NRF) of Korea funded by the Korean Government (NRF-2023R1A2C1006050). The authors used Claude (Anthropic; models Opus 5.5 and Fable 5.2, September--November 2026) to write and debug the simulation and analysis code, to carry out derivations under the authors' direction, to draft and revise the text, and to search the literature. The authors specified each calculation, checked the derivations and the numerical results on which the conclusions rest, verified every reference against its source, and take full responsibility for the content.
\end{acknowledgments}

\appendix

\section{Geometric interpretation: folded and crumpled surfaces}\label{app:geom}

\begin{figure*}[t]
\includegraphics[width=\textwidth]{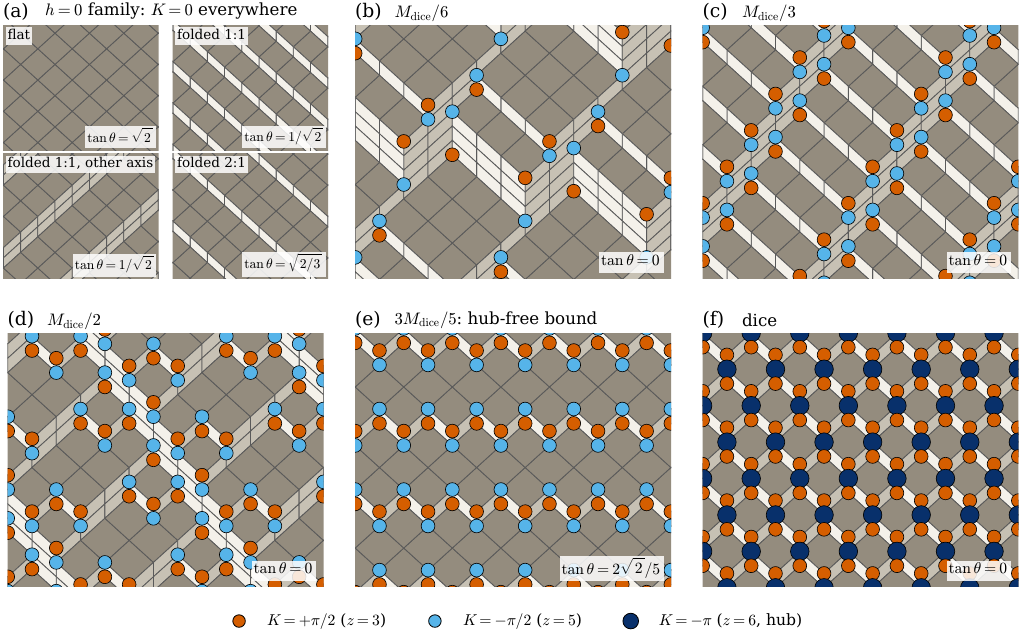}
\caption{Tilings lifted to stepped surfaces in $\mathbb Z^3$ (orthographic view, tilted by $30^\circ$ from $[111]$; faces shaded by orientation). Colored dots mark vertices with nonzero discrete Gaussian curvature $K_i=2\pi-\frac\pi2 z_i$. Each panel gives the exact tilt of the average surface from $[111]$, $\tan\theta$ (Appendix~\ref{app:geom}). (a)~Four members of the degenerate $h=0$ family, all with $K_i=0$ everywhere and all square lattices as graphs: a flat facet, and facets joined by straight folds along two different cube axes and with two different terrace ratios (the ratio of the two lozenge orientations present). (b)--(e)~Crystals of the hull at $M/\Md=\frac16,\frac13,\frac12,\frac35$. Positive curvature ($z=3$) sits only on the majority sublattice and negative curvature ($z=5$) only on the minority sublattice. The $\frac35\Md$ crystal (e) is the densest packing of sorted $\pm\pi/2$ vertices, Eq.~(\ref{eq:bound}), and the only one of (b)--(f) whose average surface is tilted. (f)~Dice: $+\pi/2$ cones and $-\pi$ hubs.}
\label{fig:curv}
\end{figure*}

\emph{Definitions.} Every lozenge tiling is the projection along $[111]$ of a stepped surface $\Sigma$ made of unit squares of the cubic lattice~\cite{Kenyon2006}. We lift the tiling by assigning the steps $+\mathbf e_x$, $+\mathbf e_y$, $+\mathbf e_z$ to the bond directions $\mathbf a_1$, $\mathbf a_2-\mathbf a_1$, $-\mathbf a_2$ (whose sum vanishes). Each lozenge becomes a unit square face, and the tiling graph becomes the edge skeleton of $\Sigma$. We verified that the lift is single-valued for all structures shown. Around a vertex of coordination $z_i$ there are exactly $z_i$ faces, each contributing a right angle, so the discrete Gaussian curvature (angle deficit) is
\begin{equation}
K_i=2\pi-\frac{\pi}{2}\,z_i=\tfrac{\pi}{2}\,(4-z_i).
\label{eq:K}
\end{equation}
Sites with $z=3,4,5,6$ are cones ($+\pi/2$), flat points ($0$), saddles ($-\pi/2$) and strong saddles ($-\pi$). Two exact identities follow. First, $\langle z\rangle=4$ is the discrete Gauss--Bonnet theorem on the torus, $\sum_iK_i=0$. Second, inserting Eq.~(\ref{eq:K}) into Eq.~(\ref{eq:charge}) gives
\begin{equation}
N_B-N_A=-\frac{1}{2\pi}\sum_i\eta_iK_i,\qquad S_{\rm tot}=\frac{S}{2\pi}\Big|\sum_i\eta_iK_i\Big| ,
\label{eq:stagK}
\end{equation}
with $\eta_i=+1$ ($-1$) on sublattice $A$ ($B$). The total curvature always vanishes, and the Lieb moment is its sublattice-staggered part. The same sum is the coefficient of the uncompensated Berry phase, $iS\sum_i\eta_i\,\omega[\mathbf n]$ for $\mathbf n_i=\eta_i\mathbf n$, which turns the two linear Goldstone modes into the single quadratic mode of a ferrimagnet (Sec.~\ref{sec:fluct}); curvature defects act as its background charge. For a region rather than the whole graph the identity holds up to a boundary term. We checked Eq.~(\ref{eq:stagK}) numerically for every structure in Fig.~\ref{fig:curv}. We distinguish this local quantity from the \emph{tilt}, the average orientation of $\Sigma$. The tilt is fixed by the fractions $f_1,f_2,f_3$ of the three lozenge orientations and is conserved by hexagon flips. The average normal of $\Sigma$ is $\propto(f_1,f_2,f_3)$, so its angle $\theta$ from $[111]$ obeys $\tan^2\theta=3(f_1^2+f_2^2+f_3^2)-1$. For a periodic tiling the $f_k$ are rational, and we quote the exact value of $\tan\theta$ rather than $\theta$.

\emph{The results in this language.} (i)~Because $z_i$ is linear in $K_i$, Eq.~(\ref{eq:coord}) states that the quantum zero-point energy is, to leading order, a convex function of the local Gaussian curvature. At zero field it is minimized by surfaces with $K_i=0$ everywhere. These are not only the flat facet (the square-type tiling) but also facets joined by straight folds along cube axes, with different fold axes and terrace widths [Fig.~\ref{fig:curv}(a)], i.e., developable, origami-like surfaces. All of them are square lattices as graphs (every vertex has degree 4 and every face is a quadrilateral), so their Hamiltonians coincide in the bulk. Consistently, the 68 distinct zero-curvature crystals (cell, orientation) in our enumeration have thermodynamic-limit LSWT energies within $2.5\times10^{-6}J$ of each other, a spread attributable to the finite $k$ grid. Folding costs nothing; curvature does. (ii)~By Eq.~(\ref{eq:stagK}) the Zeeman energy $-hS_{\rm tot}$ couples to staggered curvature. The field therefore crumples the surface, placing cones on the majority sublattice and saddles on the minority one. Every crystal of Table~\ref{tab:crystals} is \emph{perfectly sorted}, $\sum_i|K_i|=|\sum_i\eta_iK_i|$: the mean $|K_i|$ is $11^\circ$, $13^\circ$, $17^\circ$, $20^\circ$, $24^\circ$, $30^\circ$, $40^\circ$, $60^\circ$, $72^\circ$, $80^\circ$ and $120^\circ$ for $M/\Md=\frac1{11},\frac19,\frac17,\frac16,\frac15,\frac14,\frac13,\frac12,\frac35,\frac23,1$, i.e., exactly $120^\circ\times M/\Md$. No curvature is wasted on the wrong sublattice. (iii)~The hub-free kink of Sec.~\ref{sec:cost} reads: staggered curvature is cheapest in $\pm\pi/2$ vertices, and $-\pi$ saddles cost 50\% more per unit. Equation~(\ref{eq:bound}) states that $\pm\pi/2$ vertices alone can carry at most $\frac35$ of the staggered curvature of dice. The $\frac35\Md$ crystal reaches this limit, with 80\% of its vertices curved, and is the densest possible packing of perfectly sorted $\pm\pi/2$ vertices. (iv)~The field does \emph{not} control the tilt. The $\frac19\Md$, $\Md/6$, $\Md/3$, $\Md/2$ and dice surfaces all have the $[111]$ orientation, whereas the $\frac1{11}\Md$, $\frac17\Md$, $\Md/4$, $\Md/5$, $\frac35\Md$ and $2\Md/3$ crystals are tilted, with $\tan\theta=2\sqrt2/11$, $\sqrt2/7$, $\sqrt2/4$, $2\sqrt2/5$, $2\sqrt2/5$ and $\sqrt2/3$ ($\theta\approx14^\circ$, $11^\circ$, $19^\circ$, $30^\circ$, $30^\circ$ and $25^\circ$). The $h=0$ family contains orientations from $\tan\theta=1/\sqrt2$ (equal terraces) to $\tan\theta=\sqrt2$ (the flat square-type facet, a cube face) [Fig.~\ref{fig:curv}(a)]. Bending of the average surface and crumpling are distinct, and only the latter carries the moment.

\emph{Limitations of the picture.} (a)~$\Sigma$ lives in an auxiliary height space. Its third dimension encodes connectivity and is not a physical displacement. Nothing here implies that a material buckles, unless the height is itself physical, which we do not assume. (b)~The analogy with origami is exact only for the zero-field family, whose folds keep $K_i=0$. The field-induced structures contain conical defects, so they are ``crumpled with cones'' rather than isometrically folded. There are no fold angles, rigidity or mechanics: all faces are unit squares with fixed right angles. (c)~The energy is not an elastic bending energy. It is a sum of local curvature costs $\eps(z_i)$ with pair corrections between curvature defects that decay within about three lattice spacings (Sec.~\ref{sec:gas}), and there is no metric or bending modulus. The convexity is a property of the quantum spin problem, derived in Sec.~\ref{sec:ising} and confirmed by LSWT and QMC for $S=\frac12$ to $\frac32$; it is not a consequence of geometry. (d)~Equations~(\ref{eq:K}) and (\ref{eq:stagK}) and Gauss--Bonnet are exact. Statements about energies inherit the accuracy of the coordination model and of QMC discussed above. (e)~The Legendre construction that turns $e_0(m)$ into the staircase is formally the same as the construction of faceted equilibrium crystal shapes. Here, however, $m$ is a staggered curvature density and not a surface orientation, so this analogy is structural only. (f)~On the torus the tilt is fixed by the boundary conditions. The tilts quoted in (iv) are those of the optimal cells found and may be one of several nearly degenerate variants.

\section{Tests of the coordination model, the wall picture and the cooling dynamics}\label{app:tests}

\begin{figure*}[t]
\includegraphics[width=0.94\textwidth]{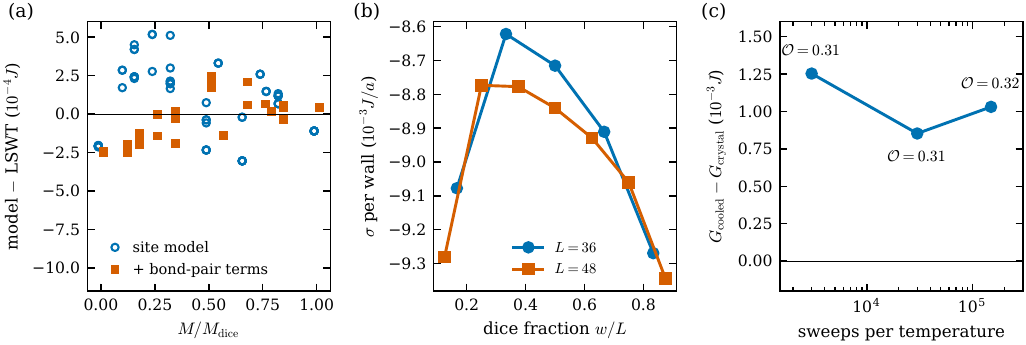}
\caption{(a)~Out-of-sample test of Eq.~(\ref{eq:coord}) on 71 periodic crystals, relative to thermodynamic-limit LSWT: site model (circles) and site model with bond-pair terms (squares). Points are offset horizontally for clarity. (b)~LSWT energy per straight dice$|$square wall for a dice strip of width $w$ in a square background on $L=36$ and $48$ tori. The walls are $w$ and $L-w$ rows apart, and closely spaced walls ($w/L$ near 0 or 1) have lower energy. (c)~Free energy $G=e_0-hm$ of the state reached by cooling to $T=0.005J$ relative to the $\Md/2$ crystal, for three cooling rates; labels give the overlap $\mathcal O$ with the crystal.}
\label{fig:A2}
\end{figure*}

\emph{Coordination model.} The training set comprises the tilt family (dice, square-type, strips of width 6 and 12, and the superlattice of alternating six-row square-type and three-row dice strips, $s6d3$) at $L=18$ and $24$, each with 0 to 600 random hexagon flips, 128 tilings in total. The site model, Eq.~(\ref{eq:coord}), gives an rms residual of $1.55\times10^{-4}J$ ($1.60\times10^{-4}J$ leave-one-out). The largest single residual is $4.1\times10^{-4}J$. Residuals averaged over each class of starting structure and over each number of flips are below $1.8\times10^{-4}J$ and show no trend with the number of flips. Adding the densities of the ten bond types $(z_i,z_j)$ reduces the rms residual to $0.82\times10^{-4}J$ ($0.87\times10^{-4}J$ leave-one-out). The out-of-sample test on 71 periodic crystals [Fig.~\ref{fig:A2}(a)] gives $2.4\times10^{-4}J$ for the site model and $1.5\times10^{-4}J$ with pair terms. Within groups of crystals that share the same $m$ and the same histogram, LSWT energies differ by up to $3.5\times10^{-4}J$ (at $\Md/3$) and $3.1\times10^{-4}J$ (at $\Md/2$). The Spearman rank correlation between the pair model and LSWT within each $m$ is $0.76$--$0.92$. For the two $\Md/2$ structures evaluated with both methods, LSWT ($-0.64945$ vs $-0.64914J$) and QMC ($\delta e=-5.3(4)$ vs $-4.8(4)\times10^{-3}J$ at $L=24$) give the same order. Two $\Md/3$ structures with the same histogram, separated by $2.5\times10^{-4}J$ in LSWT, give in QMC $-4.39(13)\times10^{-3}J$ at $L=36$ and $-3.6(3)\times10^{-3}J$ at $L=24$ (the LSWT-preferred one) and $-3.88(53)\times10^{-3}J$ at $L=18$; these values, at different sizes, do not establish the order. The best $\Md/6$ and $\Md/3$ structures of the class-resolved search lie $2.5$ and $1.1\times10^{-4}J$ below the best structures selected by histogram alone. In QMC at $L=36$ they are lower by $0.45(15)$ and $0.14(16)\times10^{-3}J$, from three and two independent runs. LSWT orderings at the level of a few $10^{-4}J$ are therefore borne out, whereas differences of order $10^{-4}J$ are not resolved by QMC.

\emph{Fitted parameters.} For $S=\frac12$ the minimum-norm least-squares solution of Eq.~(\ref{eq:coord}) on the 75 tilings of Fig.~\ref{fig:mech} is $\eps(3,4,5,6)=(-0.8544,\,-0.6582,\,-0.4423,\,-0.2061)J$ and $c_m=0.0304J$, with an rms residual of $1.4\times10^{-4}J$ compared with a spread of energies of $6.8\times10^{-3}J$. The neutral groups of Sec.~\ref{sec:cost} cost $\eps(5)+\eps(3)-2\eps(4)=0.0197J$ (a $z=3$/$z=5$ pair, $0.0099J$ per unit charge) and $\eps(6)+2\eps(3)-3\eps(4)=0.0597J$ (a hub with two $z=3$ sites, $0.0149J$ per unit charge), so the leading-order slope is $\approx0.11J$. The QMC energies follow $e_{\rm QMC}=c_0+c_1e_{\rm model}$ with $c_1=1.27$ and an rms deviation of $5$--$8\times10^{-4}J$; the QMC energy of dice lies $0.0316J$ per site above the square-lattice value. The defect lattice gas, Eq.~(\ref{eq:gas}), fitted to 4510 hub-free tilings, has $\eps_d=6.92(2)\times10^{-3}J$, $c'_m=1.99(3)\times10^{-3}J$ and bonded contacts $W=-3.24(2)\times10^{-3}J$ (like) and $+4.39(2)\times10^{-3}J$ (unlike). Its cross-validated residual is $1.5\times10^{-5}J$, compared with $2.7\times10^{-4}J$ for the site model on the same set, and $2.1\times10^{-5}J$ on 74 hub-free 48-site structures not used in the fit. Beyond $r\approx3$ the couplings are below $2\times10^{-5}J$; the moment-channel couplings at $r=2$ and $\sqrt7$ are $+0.10$ and $+0.08\times10^{-3}J$. Within each $m$ the Spearman rank correlation with LSWT is $0.93$--$0.98$.

\emph{Ising expansion.} For $S=\frac12$, Eq.~(\ref{eq:epsI}) gives $\delta^2\eps(4)=0.0095J$ and $\delta^2\eps(5)=0.0060J$, compared with about $0.020J$ from LSWT, and costs per unit moment charge of $0.0048J$ for a $z=3$/$z=5$ pair and $0.0063J$ for a hub. The contact energies of Eq.~(\ref{eq:ising}) are $+4.8\times10^{-3}J$ for an unlike contact and $-8.3$ and $-3.0\times10^{-3}J$ for $(3,3)$ and $(5,5)$ contacts. Relative to its own straight line between the square-type and dice tilings, Eq.~(\ref{eq:ising}) gives deviations of $-1.6$, $-1.9$, $-2.3$, $-2.4$, $-2.9$, $-3.2$, $-3.7$, $-4.4$, $-4.8$ and $-4.0\times10^{-3}J$ for the crystals at $M/\Md=\frac1{11},\frac19,\frac17,\frac16,\frac15,\frac14,\frac13,\frac12,\frac35$ and $\frac23$, compared with the QMC values of Table~\ref{tab:crystals}, and $e_{\rm dice}-e_{\rm sq}=0.0238J$, compared with $0.0316J$ in QMC.

\emph{Walls.} The dice$|$square wall energies extracted from the QMC strips of Fig.~\ref{fig:tilt} lie between $-0.011$ and $-0.015J/a$. In LSWT on $L=36$ and $48$ tori the energy per straight wall is $\sigma_\infty\approx-0.0088J/a$ for widely separated walls and about $0.0005J/a$ more negative when the walls are six rows apart [Fig.~\ref{fig:A2}(b)]. Straight-wall superlattices at $m=1/18$, $1/12$ and $1/9$ have LSWT energies $-0.65076$, $-0.64718$ and $-0.64246J$, compared with $-0.65281$, $-0.64945$ and $-0.64492J$ for the best crystals, i.e., $2.0$--$2.5\times10^{-3}J$ higher. Their histograms contain 6\%, 8\% and 17\% hubs, respectively.

\emph{Cooling.} Starting from the melted state at $T=0.05J$ and cooling to $0.005J$ in 19 steps with 3000, 30\,000 or 150\,000 sweeps per step ($L=24$, $h=0.19J$) always ends with $\mathcal O\approx0.31$ and $G$ above the crystal by $0.9$--$1.3\times10^{-3}J$.

\section{Numerical details, spin-wave hull and finite-size effects}\label{app:lswt}

\begin{figure}[t]
\includegraphics[width=0.9\columnwidth]{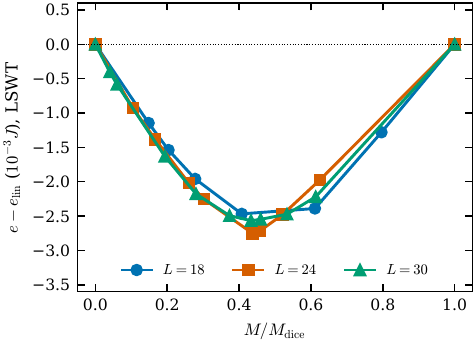}
\caption{LSWT energies of annealed tilings relative to the square-to-dice line for $L=18$, $24$, $30$. The lower envelope is size-independent to within $3\times10^{-4}J$.}
\label{fig:A1}
\end{figure}

\emph{QMC parameters and convergence.} We use $\beta J=L$ for $S=\frac12$ tori ($L=12$--$36$), $\beta J=36$ for the cell clusters, and $\beta JS=12$ for the spin-$S$ runs. The square-type tiling at $h=0$ gives $e_0=-0.66941(11)$ and $-0.66940(9)J$ at $L=36$ and $\beta J=36$ and 48, in agreement with the thermodynamic-limit value $-0.669437(5)J$~\cite{Sandvik1997}, and the $\frac35\Md$ crystal gives $\delta e=-5.43(26)$, $-5.25(11)$ and $-5.42(16)\times10^{-3}J$ at $\beta J=24$, 36 and 48, constant within errors. At the fields used ($h=0.04$--$0.24J$, below $\Delta_+$ of each crystal) the gaps to states outside the ground-state multiplet are at least $0.04J$, so that $e^{-\beta\Delta}\lesssim0.24$. Integrated autocorrelation times of the energy are $1$--$2.3$ sweeps in all runs, much shorter than the 80--150-sweep blocks used for the binning errors, and independent random seeds scatter by $0.4$--$1.4$ times the binning error. For ferrimagnetic crystals $m_{\rm meas}$ equals the Lieb value $m$ to within $10^{-4}$ in every run except the most dilute crystal. For the $\frac3{47}\Md$ crystal, simulated at $h=0.02$ and $0.04J$ because of its small gap, the components of the multiplet mix thermally and $m_{\rm meas}$ falls below $m$ by $[N(e^{\beta h}-1)]^{-1}$, as measured; $e+hm_{\rm meas}$ is unaffected, and the two fields give the same $e_0$ within errors. For the compensated square-type tiling in a field we use $e_0=e+\frac12\chi h^2$ with $\chi\simeq0.065J^{-1}$ per site, from our own runs at $L=24$ ($m/h=0.064$--$0.070$ for $h=0.1$--$0.2J$). At $L=24$ the square-type energy is converged for $\beta J\ge36$ ($-0.66967(14)$ and $-0.66950(15)J$ at $\beta J=36$ and 48); at $\beta J=24$ it lies $3\times10^{-4}J$ higher. The remaining uncertainty of the compensated reference, $\pm0.1\times10^{-3}J$, shifts the onset field by about $\pm0.006J$ but none of the steps between ferrimagnetic crystals (Sec.~\ref{sec:methods}).

\emph{Finite-size effects.} In LSWT the energy per site of the compensated square-type tiling on an $L\times L$ torus deviates from the thermodynamic limit by $-1.2\times10^{-3}$, $-1.5\times10^{-4}$ and $-4\times10^{-5}J$ at $L=12$, $24$ and $36$, whereas for the ferrimagnetic crystals the deviation is below $5\times10^{-6}J$ at $L=36$ (e.g., $-4\times10^{-8}J$ for $\Md/2$). A size-dependent term $\alpha|\mu|/N$ with $\mu=(N_A-N_B)/2$, which appears when a few uncompensated defects sit in a compensated background ($\alpha\approx3.4J$ for the staggered dice lattice), is absent for crystals with an extensive moment. The $n\times n$ cell clusters have $N=825$--$1280$ sites, and for them the LSWT energy agrees with the thermodynamic limit to $4\times10^{-5}J$ or better. In QMC the $\Md/2$ deviation is $-5.3(4)$ and $-4.98(9)\times10^{-3}J$ at $L=24$ and $36$, and $-4.8\times10^{-3}J$ at $L=12$ relative to $L=12$ references; the dice reference agrees between $L=18$ and $24$ to $10^{-4}J$.

\emph{Staggered dice lattice.} Equation~(\ref{eq:coord}) predicts $E_{\rm LSWT}/N=-0.6516J$ for the SDL, in agreement with $-0.6517(7)J$ from direct LSWT~\cite{Jagannathan2013}. The QMC energy of Ref.~\cite{Jagannathan2012}, $-0.6639J$, lies $5.5\times10^{-3}J$ above the square-lattice value. Our own SSE run, on $8\times8$ copies of the SDL cell ($N=768$, $\beta J=36$, $h=0.08J$, Zeeman energy removed with the measured moment, 900 measurement sweeps), gives $-0.6649(4)J$, i.e., $4.6(4)\times10^{-3}J$ above the square-lattice value. The $\Md/2$ crystal, with the same histogram, lies $6.2(4)\times10^{-3}J$ above our SDL value in QMC, compared with $2.3\times10^{-3}J$ in LSWT, or $2.9\times10^{-3}J$ after the linear rescaling: the cost of sublattice imbalance in QMC is about $2.7$ times its LSWT value.

\emph{Spin-wave hull.} The LSWT energies of the annealed tilings relative to the linear interpolation are independent of system size for $L=18$--$30$ (Fig.~\ref{fig:A1}). This justifies the use of $L\le36$ tori. LSWT underestimates the QMC deviations of the crystals by factors of $1.5$, $1.4$, $1.3$, $1.22$ and $1.25$ at $\Md/6$, $\Md/3$, $\Md/2$, $\frac35\Md$ and $2\Md/3$, and the step fields above $\Md/4$ by factors of $1.2$--$1.4$. For periodic crystals with an extensive moment the LSWT energy per site on an $L=36$ torus equals the thermodynamic limit to better than $5\times10^{-6}J$; no term proportional to the total moment is present (Sec.~\ref{sec:methods}). LSWT reproduces the robust part of the hull, the $\Md/3$ and $\frac35\Md$ crystals and the jump to dice, but differences of order $10^{-4}J$ are not resolved by QMC (Appendix~\ref{app:tests}). For periodic crystals we also evaluate LSWT directly in the thermodynamic limit by Bloch diagonalization of the supercell ($6\times6$ to $64\times64$ $k$ points). For the crystals checked it agrees with $L=36$ tori to $10^{-8}J$. The $\frac35\Md$ crystal is the deepest LSWT hull point as well ($-4.29\times10^{-3}J$ in the thermodynamic limit, compared with $-3.89\times10^{-3}J$ for $\Md/2$), and LSWT underestimates its QMC deviation by a factor of $1.22$. The LSWT step fields into and out of its plateau, $0.125J$ and $0.213J$, are smaller than the QMC values by factors of $1.38$ and $1.26$. In LSWT the $\frac3{47}\Md$ vertex lies only $2\times10^{-6}J$ below a $\frac1{15}\Md$ structure, at the level of the $k$-point noise, and the $\Md/2$ crystal falls on the chord between $\frac13$ and $\frac35$ to $3\times10^{-6}J$. The LSWT gaps of the moment-raising branch are $\Delta_+=0.12$, $0.15$, $0.32$, $0.43$, $0.66$, $0.82$, $0.90$ and $1.50J$ for the $\frac1{11}\Md$, $\frac19\Md$, $\Md/4$, $\Md/3$, $\Md/2$, $\frac35\Md$, $2\Md/3$ and dice crystals; the $\frac1{11}\Md$ gap ($0.120J$) coincides with the upper edge of its QMC window ($0.119J$).

\emph{Dilute limit.} In LSWT the ratio $[e_0(m)-e_{\rm sq}]/m$ is $0.0490$, $0.0492$, $0.0495$, $0.0508$ and $0.0542J$ at $M/\Md=\frac3{47},\frac1{15},\frac3{41},\frac1{11}$ and $\frac19$, so $h_c\approx0.049J$ and the magnetization reaches $\frac1{11}\Md$ within $0.007J$ of $h_c$. Applying the same saturation, scaled by the LSWT-to-QMC factor of about $1.3$, to the QMC values at $\frac3{47}\Md$, $\frac1{11}\Md$ and $\frac19\Md$ gives $0.052\pm0.009$, $0.039\pm0.007$ and $0.048\pm0.005J$, with a weighted mean $h_c=0.046\pm0.004J$, in agreement with the QMC step into $\frac1{11}\Md$ at $0.045\pm0.009J$. The QMC and LSWT onsets thus agree, whereas the QMC step fields at higher moment exceed the LSWT ones by factors of 1.2--1.4. Given the uncertainty of the compensated reference, which enters $h_c$ directly, we do not regard this difference as significant.

\emph{Spin $S$ and the $\frac35\Md$ crystal.} On clusters of 320--384 sites at $h=0$ and $\beta JS=12$, $(e_{\rm dice}-e_{\rm sq})/S=0.0626(10)$, $0.0565(15)$ and $0.0539(16)J$, $\delta e(\frac35\Md)/S=-9.2(5)$, $-8.1(5)$ and $-8.3(15)\times10^{-3}J$, and the field of the step from $\frac35\Md$ to dice is $0.257(7)$, $0.230(9)$ and $0.224(12)J$ for $S=\frac12$, 1 and $\frac32$. Linear extrapolation in $1/S$ gives $0.050J$ and $0.208J$ for $(e_{\rm dice}-e_{\rm sq})/S$ and the step field, close to the LSWT values $0.050J$ and $0.213J$.

\begin{figure}[!htbp]
\includegraphics[width=0.9\columnwidth]{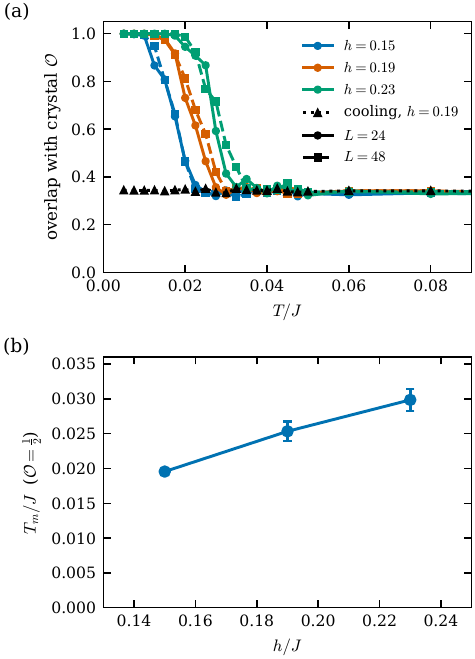}
\caption{Thermal stability of the $\Md/2$ crystal (tiling Monte Carlo with the QMC-scaled coordination model, local hexagon flips, crystal tilt sector). (a)~Overlap $\mathcal O$ of the tiling with the crystal on heating for three fields and two sizes, and on slow cooling from the disordered state (triangles). $\mathcal O\simeq\frac13$ is the value for an uncorrelated tiling. (b)~Melting temperature, defined by $\mathcal O=\frac12$.}
\label{fig:melt}
\end{figure}

\section{Finite temperature in an effective tiling model}\label{app:finiteT}

When the tiling can rearrange, for example during growth, self-assembly or in a programmable array, the plateau crystals compete with the configurational entropy of the tiling. We model this by treating the tiling as a classical degree of freedom with the coordination energy of Sec.~\ref{sec:zero}, scaled to QMC, and simulating it with local hexagon flips in the tilt sector of the crystal. We study the $\Md/2$ crystal, the deepest untilted crystal we find. The $\frac35\Md$ crystal is tilted ($\tan\theta=2\sqrt2/5$, Appendix~\ref{app:geom}) and lies in a different tilt sector, which local flips cannot reach. The spins are not sampled thermally. Because the coordination model depends only on the histogram and on $m$, all perfectly sorted arrangements with the histogram of the $\Md/2$ crystal are degenerate with it in this model. The overlap $\mathcal O$ with one representative crystal is therefore only a proxy for crystalline order, and the pair terms of the defect lattice gas (Sec.~\ref{sec:gas}) would be needed to select a single arrangement. The temperatures below are accordingly melting scales of this effective model, not transition temperatures of the full spin-plus-tiling system. Heating the $\Md/2$ crystal at fixed field [Fig.~\ref{fig:melt}(a)] produces a sharp loss of crystalline overlap at $T_m=0.020$, $0.025$ and $0.030J$ for $h=0.15$, $0.19$ and $0.23J$. The results for $L=24$ and $48$ coincide. $T_m$ increases with $h$ [Fig.~\ref{fig:melt}(b)] because the field deepens the hull minimum at $\Md/2$. In this effective model at fixed $[111]$ tilt, whose hull has only the $\Md/2$ kink, the $\Md/2$ crystal is the ground state for $h\approx0.13$--$0.25J$; all three fields lie in this window, although in the full QMC hull of Table~\ref{tab:crystals} the $\Md/2$ crystal only touches the hull. On slow cooling from the disordered state, however, the system does not recrystallize. The overlap remains at the random value and the magnetization freezes near $0.47\Md$. Cooling 50 times more slowly does not change this [Fig.~\ref{fig:A2}(c)]. The final state has $G=e_0-hm$ ($S=\frac12$) higher than the crystal by $0.9$--$1.3\times10^{-3}J$ per site. It is therefore a metastable state of the local-flip dynamics, i.e., kinetic trapping, and not evidence for an equilibrium glass phase. This trapping is a practical consideration for any realization in which the tiling is annealed.

Two omitted effects can be estimated. (i)~\emph{Magnon entropy.} From Bloch LSWT, the magnon free energy per site of the crystals at $T=0.02$--$0.03J$ is $-1.6\times10^{-6}$ (square-type) to $-7\times10^{-5}J$ (dice), roughly proportional to $m$. On melting at $h=0.19J$ the magnetization drops by $\Delta m\approx0.018$ and the configurational entropy rises by $\Delta s\approx0.1$ per site. The magnon correction shifts $T_m$ by about $\Delta F_{\rm mag}/\Delta s\sim10^{-4}J$ and the step fields by $\lesssim5\times10^{-4}J$, both negligible. (ii)~\emph{Tilt-changing moves} enlarge the configuration space of the disordered state but not of the locked crystal. They would therefore lower $T_m$, so our values are best read as upper estimates within this model. A joint treatment of spin and tiling fluctuations is left for future work.

\FloatBarrier

\end{document}